\documentclass[aps,prr,twocolumn,reprint,superscriptaddress,longbibliography]{revtex4-2}

\usepackage{hyperref}
\usepackage{graphicx}  
\usepackage{bm}        
\usepackage{amssymb}   
\usepackage{xspace}
\usepackage{verbatim}
\usepackage{color}
\usepackage{soul}
\usepackage{subfigure}
\usepackage[export]{adjustbox}
\usepackage{dcolumn}
\usepackage{amsthm,stmaryrd,mathtools}
\usepackage{txfonts}
\usepackage{braket}
\usepackage{amsmath}
\usepackage{xcolor}
\usepackage{array}
\usepackage{multirow}
\usepackage{tabularx}
\usepackage{booktabs}
\usepackage{lipsum}

\begin{document}

\author{Rozhin Yousefjani}%
\email{ryousefjani@hbku.edu.qa}
\affiliation{Qatar Center for Quantum Computing, College of Science and Engineering, Hamad Bin Khalifa University, Doha, Qatar}

\author{Shaikha Al-Naimi}%
\affiliation{Qatar Center for Quantum Computing, College of Science and Engineering, Hamad Bin Khalifa University, Doha, Qatar}

\author{Saif Al-Kuwari}%
\affiliation{Qatar Center for Quantum Computing, College of Science and Engineering, Hamad Bin Khalifa University, Doha, Qatar}

\author{Abolfazl Bayat}%
\email{abolfazl.bayat@uestc.edu.cn}
\affiliation{Institute of Fundamental and Frontier Sciences, University of Electronic Science and Technology of China, Chengdu 610051, China}
\affiliation{Key Laboratory of Quantum Physics and Photonic Quantum Information, Ministry of Education, University of Electronic Science
and Technology of China, Chengdu 611731, China}
\affiliation{Shimmer Center, Tianfu Jiangxi Laboratory, Chengdu 641419, China}

\title{Nonlinearity-enhanced Quantum Sensing in Discrete Time Crystal Probes}

\begin{abstract}
Discrete time crystals are non-equilibrium phases of matter in periodically driven systems, characterized by robust subharmonic oscillations and broken discrete time-translation symmetry. 
Their long-lived coherent dynamics and resilience to imperfections make them promising resources for quantum sensing. 
A disorder-free discrete-time crystal probe can provide the quantum-enhanced estimation of the coupling parameter. 
Here, we extend this sensing mechanism to nonlinear interactions and show that this nonlinear profile strongly enhances the sensing precision by increasing the system-size scaling exponent of the quantum Fisher information. 
Our analytical discussion separates a rigorous seminorm upper bound from the physically relevant scaling realized by product-state probes in the time crystal regime. 
Numerically, we find that the quantum Fisher information retains its quadratic long-time growth with the number of Floquet cycles, while its system-size exponent increases approximately linearly with the nonlinearity exponent, identifying nonlinearity as a resource for quantum-enhanced sensitivity.  
We further show that stronger nonlinearities shrink the time crystal stability window, making the probe more sensitive to small deviations from the resonant condition. 
We also analyze the effect of imperfect pulses and show that such imperfections can enhance, rather than suppress, the information encoded in the evolved state. 
Finally, we discuss a digital implementation of the nonlinear DTC sensing protocol using superconducting qubits.
\end{abstract}

\maketitle

\section{Introduction}

Out-of-equilibrium quantum systems can host dynamical phases that have no direct equilibrium counterpart. 
A paradigmatic example is the time crystalline phase, originally proposed as a phase of matter with spontaneous breaking of time-translation symmetry~\cite{wilczek2012quantum}. 
Although such order is forbidden in equilibrium ground states~\cite{Bruno2013b,watanabe2015absence}, periodically driven systems can evade this restriction and support discrete time crystals (DTCs), where the response of the system breaks the discrete time-translation symmetry of the drive~\cite{Sacha2015,khemani2016phase,else2016floquet,yao2017discrete,sacha2017time}.
In a period-doubled DTC, local observables oscillate with period \(2T\), even though the Hamiltonian itself is driven with period \(T\). 
The essential feature is not merely the appearance of a subharmonic response, but its rigidity. The oscillation remains locked to an integer multiple of the driving period and is robust against weak imperfections in the drive~\cite{yao2017discrete}. 
DTC order has now been observed in several platforms, including trapped ions, solid-state spin systems, ordered dipolar many-body systems, and Rydberg gases~\cite{zhang2017observation,choi2017observation,rovny2018observation,Liu2024RydbergDTC,Liu2025BifurcationTimeCrystalsRydberg,Liu2026EnhancedMultiparameterMetrology,Zhu2025DiscreteTimeQuasicrystalRydberg}. 
A central requirement for stabilizing such non-equilibrium order is to prevent the driven system from heating rapidly to a featureless infinite-temperature state. 
This can be achieved through mechanisms such as many-body localization, prethermalization, confinement, quantum many-body scars, or disorder-free localization induced by spatial gradients~\cite{alet2018many,schulz2019stark,morong2021observation,Rozha1,Rozha2,kshetrimayum2020stark,liu2023discrete,sajid2025thermal}. 
\\ \\
Many-body systems are powerful resources for quantum sensing as their collective response can amplify small parameter variations. 
In quantum metrology, this enhancement is quantified by the quantum Fisher information (QFI), whose scaling with system size determines the achievable precision. 
A prominent route to enhanced scaling is quantum criticality. 
Near a critical point, small changes in a control parameter can produce large changes in the many-body state, leading to an enhanced susceptibility and, consequently, an enhanced QFI. 
This idea has been developed in several settings, including first-order, second-order, dissipative, topological, Floquet, and Stark-localized phase transitions~\cite{raghunandan2018high,heugel2019quantum,yang2019engineering,zanardi2008quantum,invernizzi2008optimal,chu2021dynamic,montenegro2021global,fernandez2017quantum,ilias2022criticality,sarkar2022free,budich2020non,koch2022quantum,mishra2021driving,mishra2022integrable,he2023stark,yousefjani2023Long,Yousefjani2025PRApplied,yousefjani2026exponentially,gyhm2025fundamental,sarkar2025exponentially}. 
Experimental progress has also shown that critical many-body systems can be exploited for enhanced metrology in platforms such as nuclear-spin, Rydberg-atom, photonic setups, and superconducting systems~\cite{Liu2021ExperimentalCritical,Ding2022EnhancedMetrology,yu2025experimental,li2025non,Xiao2026CriticalityEnhancedQuantumSensingWalks,Xiao2024NonHermitianSensingNoEP,Tong2025TopologicalSensingQuantumWalkDefects,Beaulieu2025CriticalityEnhancedParametricResonator}. 
Despite these advances, criticality-based sensors often face practical constraints. 
They typically require the preparation of a ground state, steady state, or other specially correlated state, which can be experimentally demanding and slow. 
Moreover, the strongest enhancement is usually confined to a narrow region near the critical point, making such probes intrinsically local in parameter space. 
These limitations motivate the search for non-equilibrium sensing protocols that retain many-body enhancement while avoiding demanding state preparation.
\\ \\
DTCs provide a natural non-equilibrium route to this goal. 
Their long-lived subharmonic oscillations provide a stable dynamical reference in which precision improves quadratically with time, while the sensor exhibits strong robustness against unwanted imperfections. By increasing the number of DTC cycles, the desired signal has more time to be encoded into the system’s quantum state, thereby amplifying the amount of extractable information and enabling enhanced precision. 
This makes time-crystalline order a promising metrological resource rather than only a diagnostic of exotic dynamics~\cite{Montenegro2025Review}. 
Theoretical proposals have shown that Floquet time crystals can provide quantum-enhanced sensing of interaction parameters and periodic fields~\cite{lyu2020eternal,iemini2023floquet,yousefjani2025discrete,Yousefjani2025PeriodicFieldDTC,tsypilnikov2026exact,shukla2025prethermal}. 
Boundary time crystals and related time-crystalline phases have also been explored for continuous sensing and parameter estimation~\cite{montenegro2023quantum,Cabot2024Continuous,Viotti2026QuantumTimeCrystalClock,Gribben2025BoundaryTimeCrystalsACSensors,Sierant2022DissipativeFloquetDynamics}. 
Experimentally, DTC order has recently been used to construct highly frequency-selective sensors for AC magnetic fields~\cite{Moon2026DTCsensing}. 
A disorder-free DTC probe, as introduced in Ref.~\cite{yousefjani2025discrete}, demonstrates the advantages of this phase in sensing applications.  
It can be initialized from simple product states, does not require adiabatic ground-state preparation, and exhibits enhanced sensitivity throughout the DTC phase as well as near the DTC--non-DTC crossover. 
These features make DTC probes especially attractive for practical many-body quantum sensing.
\\ \\
In this work, we extend the DTC sensing mechanism of Ref.~\cite{yousefjani2025discrete} to nonlinear Stark-type Ising interactions. 
Specifically, we consider a periodically kicked spin chain in which the interaction strength is spatially weighted as \(j^\gamma\), where \(\gamma\) controls the degree of nonlinearity. 
This nonlinear profile modifies the parameter-imprinting generator and enhances the system-size scaling of the QFI. 
We show that increasing \(\gamma\) leads to an approximately linear growth of the QFI scaling exponent, demonstrating that nonlinear interactions provide a direct and tunable route to improved sensing precision. 
We also analyze the finite-size stability window of the DTC phase and show that stronger nonlinearities make the probe more sensitive to small deviations from the resonant condition. 
In addition, we examine the role of imperfect spin rotations and find that, for product-state initialization, a deviation from the ideal pulse can enhance the information encoded in the evolved state rather than simply degrade the sensing performance. 
Finally, we discuss a possible experimental implementation on a digital superconducting quantum simulator, where the required nonlinear interaction profile can be programmed through bond-dependent \(ZZ\) gates.

\section{Theory of quantum sensing}\label{Sec.II}

We briefly review the theory of quantum parameter estimation. 
This field aims to infer an unknown parameter $\omega$ encoded in the quantum state $\rho(\omega)$. 
The key objective is to minimize the uncertainty in estimating $\omega$, which is quantified by the standard deviation $\delta\omega$. 
This uncertainty is fundamentally bounded by the Cramér-Rao inequality~\cite{fisher1922mathematical,cramer1999mathematical}, which relates $\delta\omega$ to the classical Fisher information (CFI) $\mathcal{F}_C(\omega)$, as $\delta\omega{\geq}1/\sqrt{M\mathcal{F}_{C}(\omega)}$.
Here, $M$ represents the number of repetitions of the sensing protocol.  
For a given set of Positive Operator-Valued Measurement (POVM) operators $\{\Pi_r\}$, any measurement outcome $r$ appears with the probability $p_r(\omega){=}{\rm Tr} [\rho(\omega)\Pi_r]$. One can obtain the CFI as $\mathcal{F}_{C}(\omega){=}\sum_{r}p_r(\omega) [\partial_\omega\ln p_r(\omega)]^2$, with $\partial_\omega$ being the derivatives with respect to the parameter $\omega$.
Optimizing over all possible POVMs $\{\Pi_r\}$ leads to the quantum Cramér-Rao inequality, which is bounded by the QFI $\mathcal{F}_Q(\omega)$
as 
\begin{equation}\label{Eq.QCR}
\delta\omega{\geq}\frac{1}{\sqrt{M\mathcal{F}_{C}(\omega)}} {\geq}\frac{1}{\sqrt{M\mathcal{F}_{Q}(\omega)}}.  
\end{equation}
The QFI represents the ultimate precision limit in quantum sensing.
For pure quantum states $\rho(\omega) {=} |\psi(\omega)\rangle\langle\psi(\omega)|$, the QFI takes the standard form~\cite{braunstein1994statistical}
\begin{equation}\label{Eq.QFI}
\mathcal{F}_{Q}(\omega){=}4 \left(\langle \partial_{\omega}\psi(\omega) | \partial_{\omega}\psi(\omega) \rangle{-}|\langle \partial_{\omega}\psi(\omega) | \psi(\omega) \rangle|^2 \right).    
\end{equation}
The performance of a sensor is characterized by how the QFI scales with available resources, such as probe size or time. Specifically, this relationship can be expressed as $\mathcal{F}_Q {\sim}L^\beta$, where $L$
represents the resource and $\beta$ is the corresponding exponent.
In the absence of quantum effects, the QFI can at best scale linearly with $L$, i.e. $\beta{=}1$, known as the standard limit. 
However, by exploiting quantum probes with quantum features, it is possible to enhance the precision to $\beta{>}1$, a phenomenon referred to as quantum-enhanced sensitivity. 
A notable case is known as Heisenberg scaling in which $\beta{=}2$. 
Protocols using maximally entangled states, like Greenberger–Horne–Zeilinger (GHZ) states, are well-known examples of achieving Heisenberg scaling.
Strongly correlated many-body systems can go beyond Heisenberg scaling, with $\beta{>}2$, by harnessing a variety of quantum features such as criticality. Nonetheless, the obtainable precision is fundamentally bounded by the details of the interaction in the system~\cite{Boixo2007GeneralizedLimits,Abiuso2025FundamentalLimitsThermalEquilibrium,Puig2025DynamicalSteadyStateManyBodyMetrology,Montenegro2025Review}.

\section{Model}\label{Sec.III}
Consider a one-dimensional chain containing \(L\) spin-\(1/2\) particles with a nonlinear Stark-type Ising interaction, periodically kicked by a global spin rotation. 
The Hamiltonian reads
\begin{align}\label{Eq.Hamiltonian}
H(t) = J H_I^{(\gamma)} + \sum_n \delta(t-nT)H_P, 
\\
H_I^{(\gamma)} = \sum_{j=1}^{L-1}j^{\gamma}\sigma^{z}_{j}\sigma^{z}_{j+1}, 
\qquad  
H_P  = \Phi \sum_{j=1}^{L}\sigma^{x}_{j}.
\end{align}
Here, \(J\) is the spin-exchange coupling, \(T\) is the driving period, \(\sigma^{x,y,z}_{j}\) are the Pauli operators, and \(\Phi{=}(1{-}\varepsilon)\pi/2\) is the pulse angle, with \(\varepsilon\) quantifying the error from an ideal spin flip.  
The Stark-type \(zz\) interaction is linear for \(\gamma{=}1\) and nonlinear for integer \(\gamma>1\).
Regardless of whether it is linear or nonlinear, this interaction generates off-resonant energy splitting at each site, causing the particle's wave function to localize~\cite{Rozha1,Rozha2}. 
This effect is similar to the localization typically produced by applying a gradient magnetic field~\cite{schulz2019stark,morong2021observation,he2023stark,yousefjani2023Long,Yousefjani2025PRApplied} and is essential to prevent our system from absorbing the energy of periodic drives~\cite{lenarvcivc2020critical,Rozha2}.
The absence of this mechanism causes an unbounded increase in entropy and thermalization of the system~\cite{d2014long,lazarides2014equilibrium}.

Since \(H_P\) acts as a periodic kick, the single-cycle Floquet operator is 
\begin{equation}\label{Eq.FloquetUnitary}
U_{F}(T,\varepsilon) = e^{-iH_{P}} e^{-iJT H_I^{(\gamma)}}.    
\end{equation}
In order to keep the presentation self-contained, it is useful to briefly recap the obtained results in Ref.~\cite{yousefjani2025discrete} regarding the linear gradient interaction ($\gamma{=}1$). 
Then we address the effect of the nonlinear gradient interaction on the coherent behavior of the system.
In Ref.~\cite{yousefjani2025discrete} it has been shown that in a system with linear interaction ($\gamma{=}1$) and fixed $JT{=}\pi/2$, a stable period-doubling DTC occurs for arbitrary pulse error $\varepsilon$ and computational-basis initial states.  
Therefore observing the evolution of a system that is initialized in an arbitrary computational basis $|\mathbf{z}\rangle$  
over $n$ period cycles using the revival fidelity $F(nT){=}|\langle \mathbf{z}|U_F^{n}(JT{=}\pi/2,\varepsilon{\neq}0)|\mathbf{z}\rangle|^2$
demonstrates that the system returns to its initial state after $2T$, namely $F(2nT){=}1$.
These coherent oscillations of $F(nT)$ in the DTC phase show infinite persistence even in a system with finite sizes.
Interestingly, the nonlinearity of the interaction does not alter this resonant period-doubling mechanism. 
The operator \(H_I^{(\gamma)}\) is diagonal in the computational basis \(\{|\mathbf z\rangle\}\) and is invariant under the global spin flip \(X=\prod_j\sigma_j^x\), namely \([H_I^{(\gamma)},X]=0\). 
Therefore, each configuration and its globally flipped partner have the same interaction energy. 
For integer \(\gamma\), the resonant phases generated at \(JT=\pi/2\) have the same parity structure as in the linear case, since \(j^\gamma\) and \(j\) have the same parity. 
Consequently, the derivation of Ref.~\cite{yousefjani2025discrete} carries over to nonlinear Stark profiles, and tuning \(JT=\pi/2\) produces an eternal period-doubled response.

\section{Nonlinear DTC probe: analytical analysis}\label{Sec.DTC_Probe}
Most of the previous proposals on DTC are stable within a range of $JT$.
In contrast, our period-doubling DTC exhibits extreme sensitivity to $JT$.
This reveals its potential to sense either \(J\) or \(T\).
In Ref.~\cite{yousefjani2025discrete} the capability of this system to measure the signed detuning $\omega{=}\pi/2{-}JT$ has been investigated. 
It turns out that only when the distance from resonance is smaller than a specific value denoted by $\omega_{\max}$, one has a stable DTC order. 
In fact, coherent oscillations of the revival fidelity $F(nT)$ for $\omega{<}\omega_{\max}$ shift to nontrivial oscillations for $\omega{>}\omega_{\max}$, indicating a finite-size transition-like crossover from DTC order to a non-DTC region.
This crossover scale is nearly unaffected by $\varepsilon$ and is observable at all stroboscopic times $n$. 
Notably, the transition scale is connected to the size of the probe and extracted as $\omega_{\max}\propto L^{-1}$, showing a gradual shift towards smaller values as the probe size increases.
It has been shown that the tiny shift $\omega$ in this DTC can be estimated by precision $\mathcal{F}_{Q}(\omega)\propto n^{\alpha}L^{\beta}$ 
with $\alpha{=}2$ in the long term and $\beta{\cong}3$ in both DTC phase and at the finite-size threshold.

In what follows, we aim to analyze the effect of nonlinear interaction on the precision limit in sensing $\omega$. For calculating the QFI for an evolved pure state after \(n\) Floquet cycles, namely $|\psi_n(\omega)\rangle {=} U_\omega^n|\psi_0\rangle$, one needs to obtain $\partial_\omega |\psi_n(\omega)\rangle = i\mathcal G_n |\psi_n(\omega)\rangle$ wherein $\mathcal G_n$ is the effective parameter-imprinting generator.  
Starting from $U_\omega {=} e^{-iH_P} e^{-i(\pi/2-\omega)H_I^{(\gamma)}}$ in which  $H_I^{(\gamma)} {=} \sum_{j=1}^{L-1} j^\gamma \sigma_j^z\sigma_{j+1}^z$, this generator can be obtained as 
$\mathcal G_n = \sum_{m=0}^{n-1} U_\omega^m \widetilde H_I^{(\gamma)} U_\omega^{-m}$ where $\widetilde H_I^{(\gamma)} = e^{-iH_P}H_I^{(\gamma)}e^{iH_P}$
is the pulse-dressed interaction Hamiltonian.
This operator is Hermitian and has the same spectrum as \(H_I^{(\gamma)}\).
The QFI is therefore exactly 
$\mathcal F_Q(\omega) = 4\,{\rm Var}_{\psi_n}(\mathcal G_n) = 4 \left( \langle \mathcal G_n^2\rangle - \langle \mathcal G_n\rangle^2 \right).$
Thus, the scaling of the sensing precision is controlled by the variance of \(\mathcal G_n\). A rigorous upper bound follows directly from the seminorm of \(\mathcal G_n\). 
For any Hermitian operator \(A\), ${\rm Var}_\psi(A) \leq \frac{1}{4}\|A\|_s^2 ,$ where $\|A\|_s = \lambda_{\max}(A)-\lambda_{\min}(A)$ is the spectral seminorm.
This inequality is saturated only if the state $\psi$ has equal probability weight on the maximum- and minimum-eigenvalue subspaces of $A$.
Considering this upper bound, one has $\mathcal F_Q(\omega) \leq \|\mathcal G_n\|_s^2$.
The seminorm is invariant under unitary transformations and satisfies the triangle inequality, which results in
\begin{align}
\|\mathcal G_n\|_s
\leq
\sum_{m=0}^{n-1}
\left\|
U_\omega^m
\widetilde H_I^{(\gamma)}
U_\omega^{-m}
\right\|_s
=
n\|H_I^{(\gamma)}\|_s.
\label{eq:Gn_seminorm}
\end{align}
Because all terms in \(H_I^{(\gamma)}\) commute, one has  $\lambda_{\max} {=} \sum_{j=1}^{L-1}j^\gamma$, $\lambda_{\min} {=} -\sum_{j=1}^{L-1}j^\gamma$ as the largest and smallest eigenvalues of $H_I^{(\gamma)}$.
Hence, $\mathcal F_Q(\omega) \le 4 n^2 \left( \sum_{j=1}^{L-1}j^\gamma \right)^2 .$
Using the large-\(L\) asymptotic form, one has 
\begin{equation}
\mathcal F_Q(\omega)
\lesssim
n^2 L^{2\gamma+2}.
\label{eq:seminorm_scaling}
\end{equation}
This is a rigorous ceiling on the attainable QFI. 
To saturate the first inequality, namely $\mathcal F_Q(\omega) \leq \|\mathcal G_n\|_s^2$, the final probe state needs to have an equal-weight superposition of eigenstates corresponding to the largest and smallest eigenstates of $\mathcal G_n$.
Each term $A_m:=U_\omega^m \widetilde H_I^{(\gamma)} U_\omega^{-m}$ in $\mathcal G_n$ has the same spectrum as $H_I^{(\gamma)}$, because it is related to this operator by unitary transformations. However, the different $A_m$'s do not generally commute and hence do not share the common extremal eigenstates, necessary to satisfy the triangle inequality. Therefore, the eigenstates of $\mathcal G_n = \sum_{m=0}^{n-1} A_m$ are not generally the same as the eigenstates of $H_I^{(\gamma)}$.
So, for generic $\omega$ and $\varepsilon$, the largest and smallest eigenstates of $\mathcal G_n$ are generally Floquet-dressed many-body entangled states, not simply computational-basis product states.
The perfect-pulse limit is instructive. 
For \(\varepsilon=0\), the pulse is a global spin flip, and therefore \(\widetilde H_I^{(\gamma)}=H_I^{(\gamma)}\). 
Moreover, \([U_\omega,H_I^{(\gamma)}]=0\) for any detuning \(\omega\), so that \(\mathcal G_n=nH_I^{(\gamma)}\) exactly. 
In this limit, the largest-eigenvalue subspace is spanned by \(\{|{\uparrow \uparrow \cdots \uparrow} \rangle, |{\downarrow \downarrow \cdots \downarrow} \rangle\}\), while the smallest-eigenvalue subspace is spanned by the two Néel configurations \(\{|{\uparrow \downarrow \uparrow \downarrow \cdots} \rangle, |{\downarrow \uparrow \downarrow \uparrow\cdots}\rangle\}\). 
The seminorm bound can be saturated only by a final state with equal weight on these extremal subspaces. 
By contrast, the ferromagnetic product state used below is itself an eigenstate of \(H_I^{(\gamma)}\) hence it has zero variance and zero QFI in the ideal-pulse limit. 
Thus, for product-state initialization, a finite pulse imperfection is not merely a nuisance but is the mechanism that mixes different interaction-energy sectors and allows the detuning to be imprinted on the state.

\section{Nonlinear DTC probe: numerical analysis}
In the following, we numerically show that, for imperfect pulse rotations, namely \(\varepsilon{\neq}0\), a simple product-state initialization realizes the scaling 
\begin{equation}
\mathcal F_Q(\omega)
\propto 
n^\alpha L^{\beta}, \quad \mathrm{with} \quad \alpha\simeq2, \quad \beta \simeq 2\gamma+1.
\label{eq:uni_scaling}
\end{equation} 
First, we address the case $\gamma{=}2$, then complete the study by considering other values of $\gamma$ and extracting the universal scaling formula in Eq.~(\ref{eq:uni_scaling}).

\subsection{Case $\gamma=2$}\label{Sec.gamma2}
While our results are generic and remain valid for all choices of computational basis states as the initial state, we focus on $|\psi_{0}\rangle{=}|\mathbf{0}\rangle{=}|{\uparrow,\cdots,\uparrow}\rangle$.  
In this paper, the results for $L{\leqslant}14$ have been obtained using exact diagonalization (ED) and for larger system sizes we use time-dependent variational principle (TDVP) techniques for finite matrix product state (MPS), using PYTHON package TeNPy~\cite{tenpy}. 
\begin{figure}
    \centering
    \includegraphics[width=0.49\linewidth]{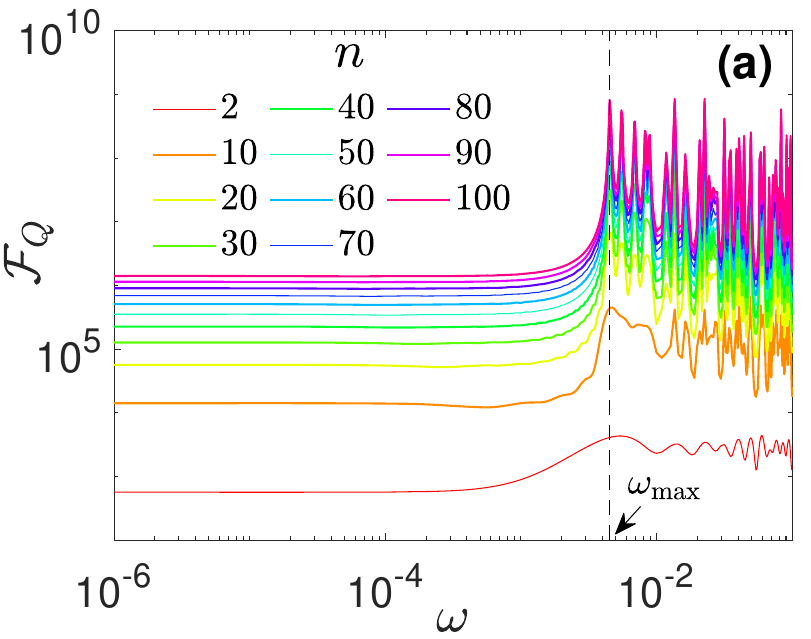}
    \includegraphics[width=0.49\linewidth]{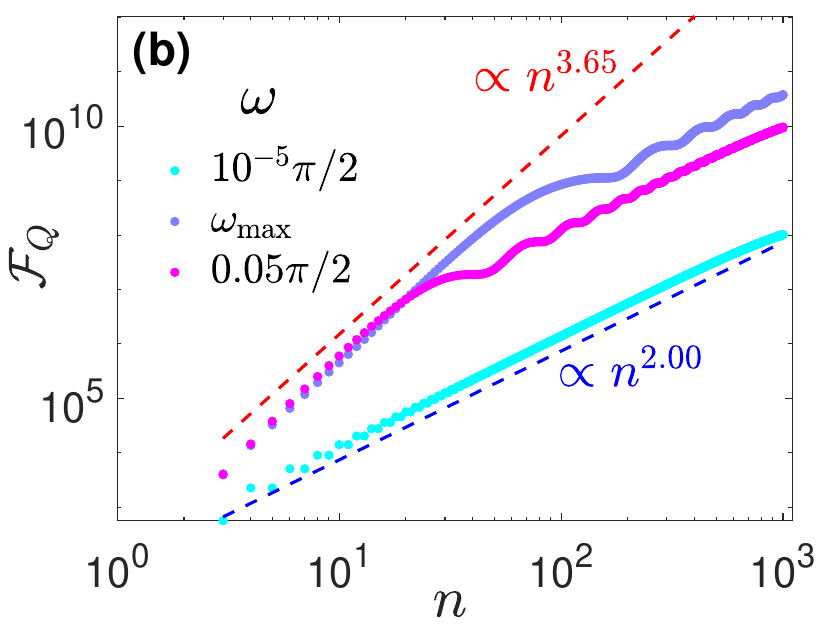}
        \caption{(a) The evolution of the QFI as a function of the deviation $\omega$ over various period cycles $n$, in a DTC probe with size $L{=}12$, fixed pulse error $\varepsilon{=}0.01$ in the $\pi/2$ $x$-rotation, and with the nonlinearity degree $\gamma{=}2$. The dashed line marks the finite-size threshold $\omega_{\max}$ separating the DTC phase and non-DTC region. (b) The dynamics of the QFI over thousands of period cycles obtained for a DTC probe operating within the DTC phase for $\omega{=}10^{-5}\pi/2$, at the finite-size threshold $\omega_{\max}$, and inside the non-DTC region for $\omega=0.05\pi{/}2$. The other parameters are similar to the panel (a).   }
    \label{fig:1}
\end{figure}

In Fig.~\ref{fig:1} (a) we plot the QFI  $\mathcal{F}_{Q}$ as a function of $\omega$ at different stroboscopic times $n{\in}\{2,10,\cdots,100\}$, in a chain of length $L{=}12$ and $\varepsilon{=}0.01$. 
Several interesting characteristics are illustrated here. 
First, the QFI exhibits unique behaviors within each phase: during the DTC phase, namely for $\omega{<}\omega_{\max}$, the QFI stabilizes into a plateau whose magnitude is $n$ dependent, whereas in the non-DTC sector, determined by $\omega{>}\omega_{\max}$, it demonstrates nontrivial and rapid oscillations. 
Second, as the system approaches the finite-size threshold, denoted by $\omega_{\max}$  and indicated by a dashed line, the QFI prominently displays a distinct peak at every stroboscopic time.
Third, while the peaks get sharper by increasing $n$, its position $\omega_{\max}$  does not depend on $n$.
To explore the evolving behavior of the QFI, in Fig.~\ref{fig:1} (b)
we plot $\mathcal{F}_{Q}$ over thousands of driving cycle $n$ in a system of size $L{=}12$ and $\varepsilon{=}0.01$ at three different $\omega$'s, namely deep inside the DTC phase for $\omega{=}10^{-5}\frac{\pi}{2}$, in the finite-size threshold $\omega_{\max}$, and in the non-DTC region for $\omega{=}0.05\frac{\pi}{2}$.
Deep inside the DTC phase one observes the quadratic scaling of the QFI, $\mathcal{F}_{Q}\propto n^2$. 
For other values of the deviation $\omega$, one notices a rapid growth in early times as $\mathcal{F}_{Q}\propto n^{3.65}$ that after a transient time turns to quadratic scaling.
Interestingly, the time interval for this rapid growth of the QFI in the finite-size threshold is considerably long. 
In most of the sensing setups, there are often time constraints for completing the sensing process. 
Hence, the accelerated algebraic growth of the QFI beyond quadratic can lead to a significant enhancement in measurement precision.
Obviously, the DTC probe, tuned in the vicinity of the finite-size threshold can yield higher precision in sensing $\omega$ even in early period cycles.
These results are in line with our previous results for the linear DTC sensor, namely $\gamma{=}1$, presented in Ref.~\cite{yousefjani2025discrete}.
\begin{figure}
    \centering
    \includegraphics[width=0.49\linewidth]{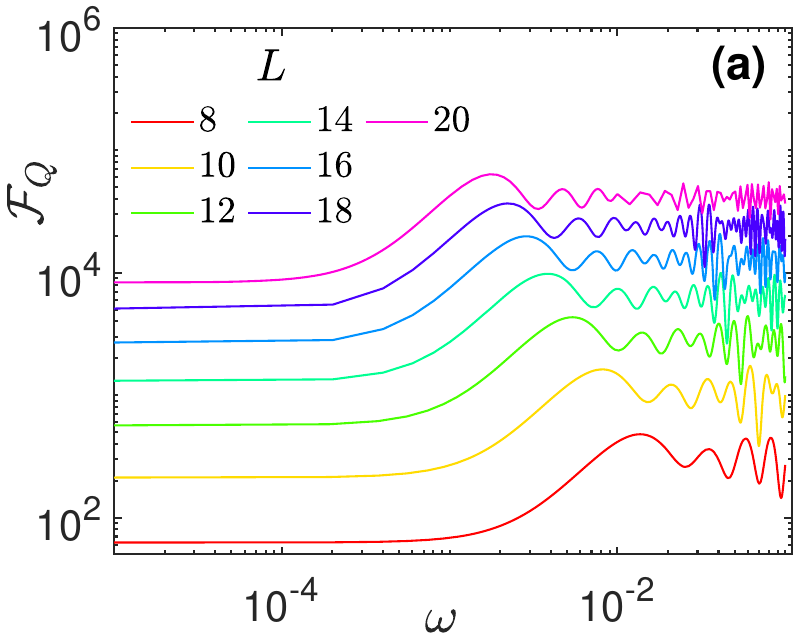}
    \includegraphics[width=0.49\linewidth]{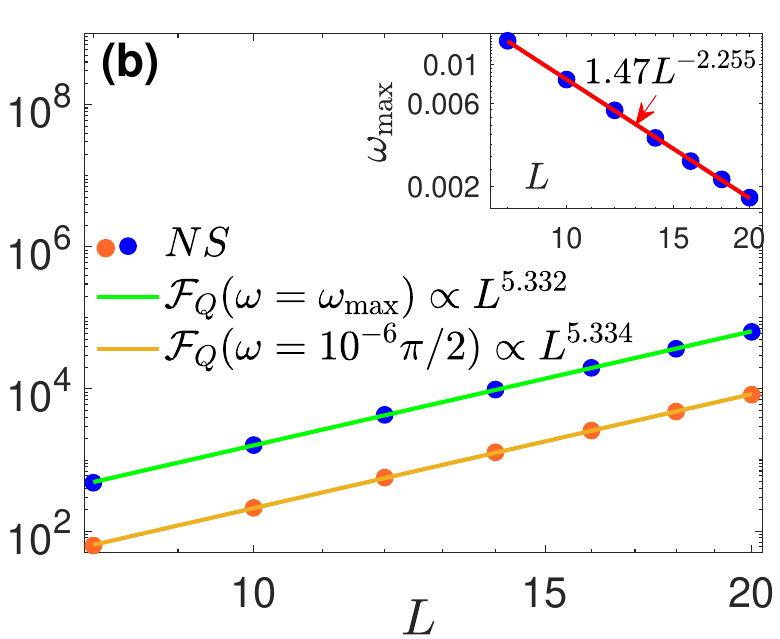}
        \caption{(a) The QFI versus the deviation $\omega$ in DTC probe with various sizes, fix $\varepsilon{=}0.01$, and nonlinearity degree $\gamma{=}2$. The results are obtained after $n{=}2$ period cycle.  (b) The extracted QFI for two determined values of the $\omega$ as a function of the system size $L$. The numerical simulations (NS) are well described by the fitting function $\mathcal{F}_{Q}\propto L^{\beta}$. 
        For a DTC probe with $\omega{=}10^{-6}\pi/2$ namely within the DTC phase we achieve $\beta{=}5.334 $. At the finite-size threshold $\omega_{\max}$ the obtained exponent is $\beta{=}5.332$. The inset depicts the extracted $\omega_{\max}$ at different $L$'s. The numerical results (dots) follow the algebraic function $\omega_{\max}\propto L^{-2.255}$.
        }
    \label{fig:2}
\end{figure}

To identify more possibilities of quantum-enhanced precision, we analyze the QFI at various system sizes $L{\in}[8,\cdots,20]$. 
In Fig.~\ref{fig:2} (a) we depict the QFI as a function of $\omega$, with the chain size varying while $\varepsilon{=}0.01$ remains constant.
Since the transition between two DTC and non-DTC phases is evident across all stroboscopic times, as shown in Fig.~\ref{fig:1} (a), to extract the scaling behavior of the QFI, attention should be directed to a specific period cycle. 
This allows us to discriminate the enhancement caused by system size from period cycles. Here we select $n{=}2$.
One can observe the finite-size scaling in both the DTC phase and the finite-size threshold. 
Fig.~\ref{fig:2} (b), displays the extracted QFI as a function of system size $L$ deep inside the DTC phase for  $\omega{=}10^{-6}\pi/2$, and also at the finite-size threshold $\omega{=}\omega_{\max}$.
The numerical results can be appropriately represented by a fitting function of the form $\mathcal{F}_{Q}\propto L^{\beta}$ with $\beta{=}5.334$ in the DTC phase and $\beta{=}5.332$ at the finite-size threshold.
Several notes need to be highlighted. 
As has been mentioned before, using classical probes at best can result in $\beta{=}1$.  
However, by leveraging quantum properties, precision can be heightened beyond $\beta{>}1$.
The obtained results evidence that our DTC is indeed capable of delivering quantum-enhanced sensitivity when it is used for measuring the deviation $\omega$. 
Note that for linear interaction  $\gamma{=}1$, one obtains  $\beta{\sim}3$ inside the DTC phase and also at the finite-size threshold~\cite{yousefjani2025discrete}.
The enhancement of exponent $\beta$ by increasing nonlinearity $\gamma$ clearly shows the improvement of the DTC probe through harnessing nonlinearity. 
In Fig.~\ref{fig:2} (a), as the chain size increases, the peaks of the QFI gradually shift towards lower values of $\omega$. 
In the inset of Fig.~\ref{fig:2} (b), we plot the corresponding $\omega_{\max}$ as a function of system size $L$. 
The numerical results are well-mapped by the function $\omega_{\max}{\cong}1.47 L^{-2.255}$. 
This shows that adding nonlinearity to the interaction makes the DTC probe more susceptible to the deviation $\omega$.

\begin{figure}
    \centering
    \includegraphics[width=0.49\linewidth]{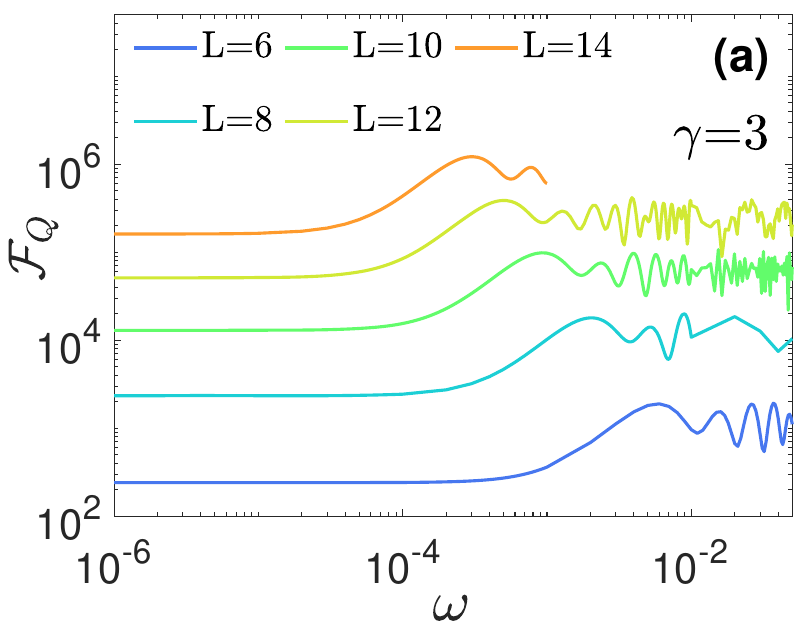}
    \includegraphics[width=0.49\linewidth]{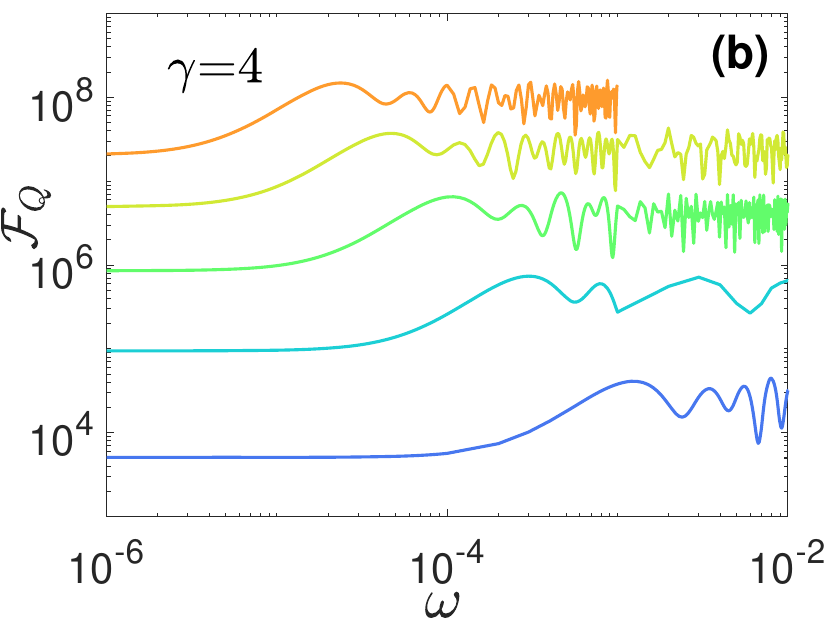}
    \includegraphics[width=0.49\linewidth]{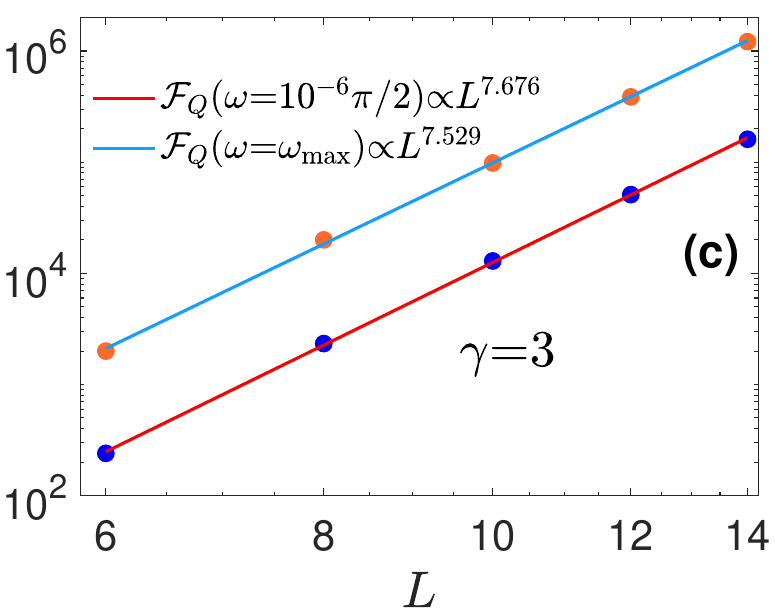}
    \includegraphics[width=0.49\linewidth]{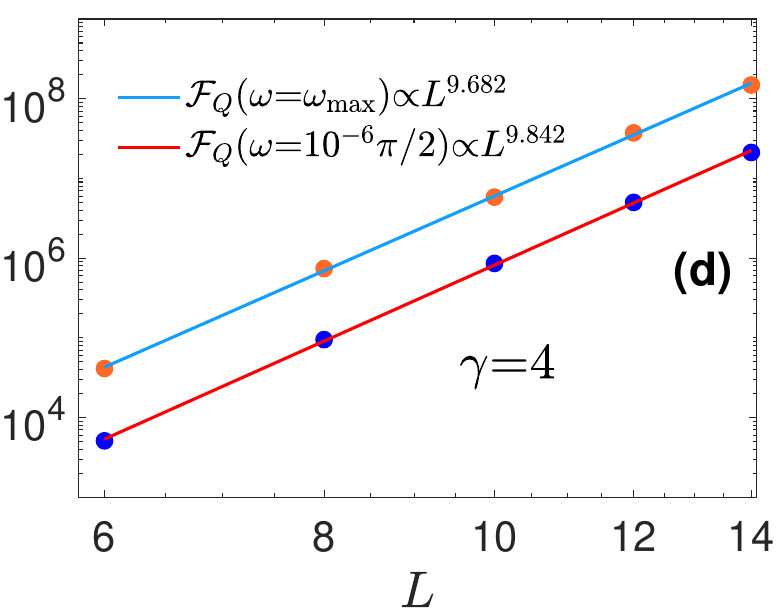}
    \includegraphics[width=0.49\linewidth]{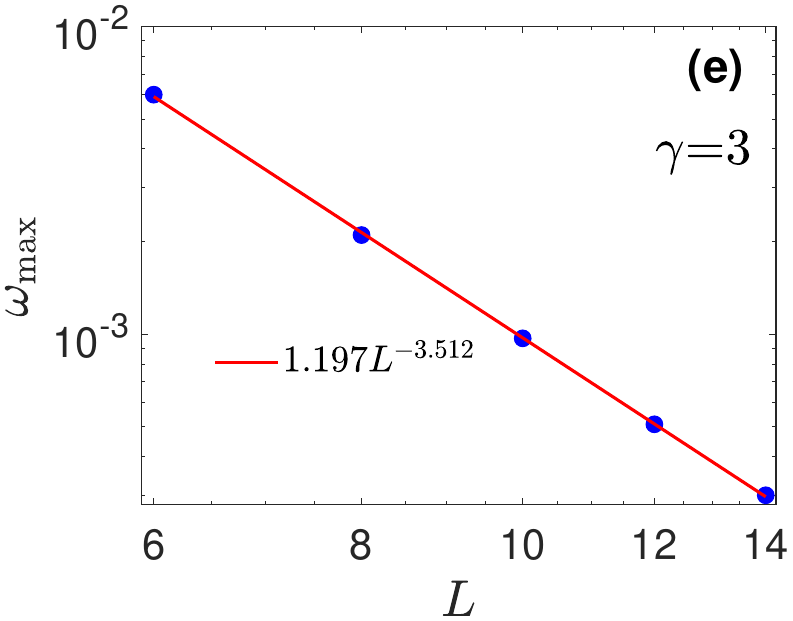}
    \includegraphics[width=0.49\linewidth]{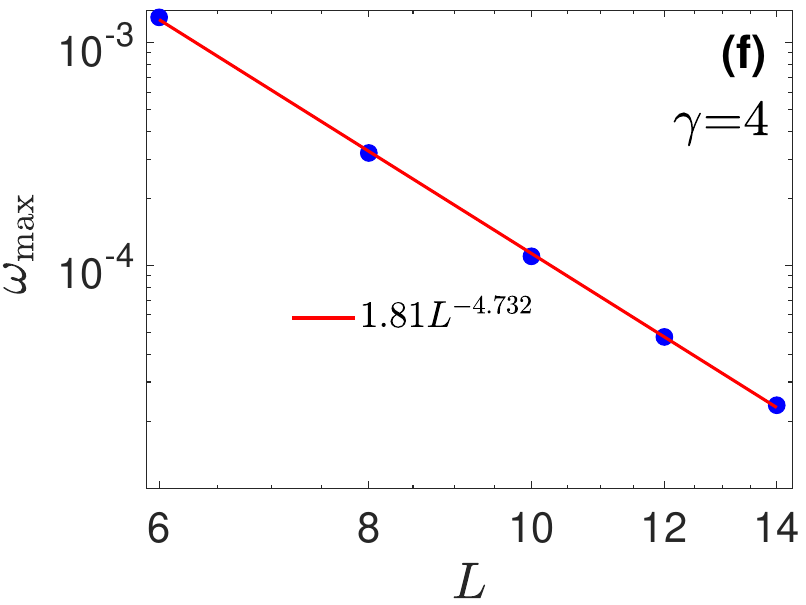}
    \caption{The QFI as a function of the deviation $\omega$ over $n=2$ period cycles in DTC probes with (a) $\gamma{=}3$ and (b) $\gamma{=}4$. The results in both panels are obtained for different system sizes but fixed $\varepsilon{=}0.01$. The obtained QFI within the DTC phase for $\omega{=}10^{-6}\pi/2$ and at the finite-size threshold $\omega_{\max}$ in DTC probes with (c) $\gamma{=}3$ and (d) $\gamma{=}4$. The numerical results (dots) in both panels are well-mapped by the fitting function $\mathcal{F}_{Q}\propto L^{\beta}$ for the reported $\beta$'s in the corresponding plots. The extracted finite-size threshold \(\omega_{\max}\) is shown in panels (e) and (f) for \(\gamma=3\) and \(\gamma=4\), respectively. The numerical results (dots) are properly fitted by the functions $\omega_{\max}\propto L^{-3.512}$ and $\omega_{\max}\propto L^{-4.732}$, for $\gamma{=}3$ and $\gamma{=}4$, respectively.  }
    \label{fig:3}
\end{figure}

\subsection{Case $\gamma>2$}\label{Sec.V}

Having elucidated the effect of nonlinearity with $\gamma{=}2$ on the sensing precision, we now aim to establish a universal algebraic formula that can properly predict the scaling of the QFI concerning $\gamma$. To achieve this, we extend our study to $\gamma{=}3$ and $4$. 
In Fig.~\ref{fig:3} (a)-(b), we plot the QFI as a function of the deviation $\omega$ after $n{=}2$ period cycle in chains with various system sizes and nonlinearity degrees $\gamma{=}3$ and $\gamma{=}4$, respectively.
Note that results are obtained for $\varepsilon{=}0.01$. 
Analogs to the case $\gamma{=}2$, we investigate the scaling behavior of the QFI both within the DTC phase, for $\omega{=}10^{-6}\pi/2$, and at the finite-size threshold $\omega_{\max}$. 
Fig.~\ref{fig:3} (c)-(d) depict the corresponding values for the QFI as a function of $L$ for the nonlinearity $\gamma{=}3$ and $\gamma{=}4$, respectively.
In both cases, numerical results (dots) align well with the fitting function as $\mathcal{F}_{Q}\propto L^{\beta}$.
For $\gamma{=}3$, we observe $\beta{=}7.676$ inside the DTC phase and $\beta{=}7.529$ at the finite-size threshold. For $\gamma{=}4$, the respective exponents are $\beta{=}9.842$ and $\beta{=}9.982$ within the DTC phase and at the finite-size threshold, respectively.
Results from Ref.~\cite{yousefjani2025discrete} for $\gamma{=}1$ combined with the preceding analysis, propose the following ansatz for the QFI  
\begin{equation}
\mathcal{F}_{Q}\propto n^2 L^{\beta},  \quad \mathrm{with} \quad  \beta = a\gamma +b 
\end{equation}
with $(a,b){=}(2.249,0.871)$ and $(a,b){=}(2.236,1.062)$ for a DTC probe operating inside the DTC phase and at the finite-size threshold, respectively.
This universal formula serves as the primary outcome of this study, elucidating the relationship between degrees of nonlinearity and the precision scaling of the QFI. 
Focusing on the finite-size threshold, we plot the extracted $\omega_{\max}$ for both DTC probes with nonlinearity $\gamma{=}3$ and $\gamma{=}4$ in Fig.~\ref{fig:3} (e)-(f). 
The fitting function properly describes the numerical results (dots) as $\omega_{\max}{\cong}1.197 L^{-3.512}$ and $\omega_{\max}{\cong} 1.81L^{-4.732}$. 
Regarding the extracted $\omega_{\max}$ for various $\gamma$'s
we obtain
\begin{equation}
    \omega_{\max} \propto L^{-1.265 \gamma}.
\end{equation}
As the system size increases, the transition between DTC phase and the region with unbroken time translational symmetry takes place for smaller values of deviation. This algebraic decrease indicates that nonlinearity can increase the susceptibility of the DTC probe to the driving frequency.  
\\ \\
Up to now, we have focused on estimating the dimensionless detuning $\omega=\left|\pi/2-JT\right|$
which measures the distance from the resonant DTC condition.
This parameter can also be interpreted as a frequency detuning when the coupling \(J\) is fixed and the pulse period \(T\) is controlled by an external drive. 
Let the pulses be generated by a time-periodic field producing a global spin rotation about the \(x\)-axis, with repetition period \(T=2\pi/\omega_d\), where \(\omega_d\) is the angular driving frequency. 
Then
\(
\omega= \left| \pi/2 - {2\pi J/\omega_d} \right|.
\)
The exact DTC resonance occurs at $\omega_d=4J $.
Thus, the DTC probe is frequency selective around \(\omega_d=4J\). 
In a finite chain, the DTC response remains stable only within the window
$\left|
\pi/2
-
{2\pi J/\omega_d}
\right|
<
\omega_{\max}(L),$
or, equivalently near resonance $|\omega_d-4J|
\lesssim \frac{8J}{\pi}\omega_{\max}(L)$.
The QFI scaling obtained for the detuning \(\omega\), \(\mathcal F_Q(\omega)\propto n^2L^\beta\), therefore translates into the same scaling for estimating \(\omega_d\), up to the constant factor
\[
\mathcal F_Q(\omega_d)
=
\mathcal F_Q(\omega)
\left(
\frac{\partial \omega}{\partial \omega_d}
\right)^2
\simeq
\mathcal F_Q(\omega)
\left(
\frac{\pi}{8J}
\right)^2 .
\]
Consequently, the standard deviation of the frequency estimate scales as
\[
\delta\omega_d
\gtrsim
\frac{8J}{\pi\sqrt{M}}\,
n^{-1}L^{-\beta/2}.
\]
\begin{figure}[t!]
    \centering
    \includegraphics[width=0.49\linewidth]{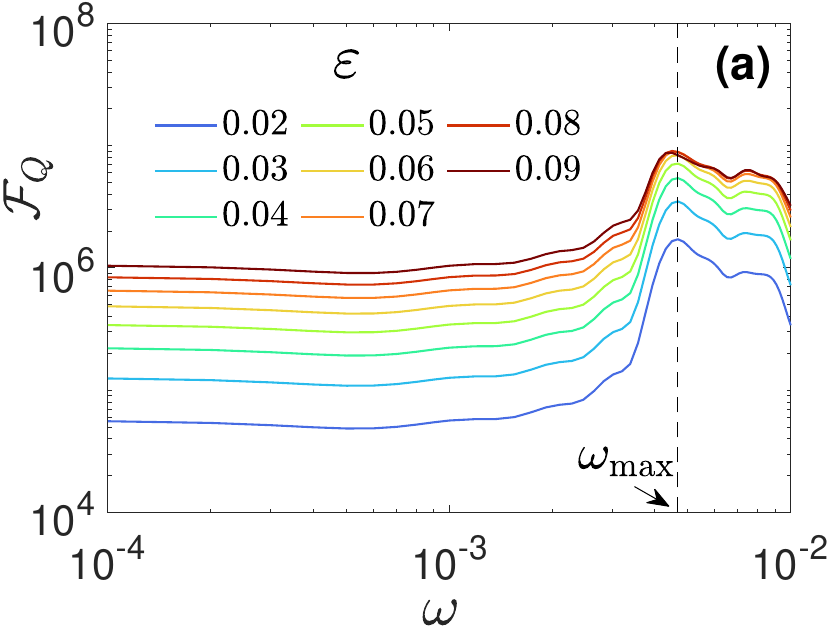}
    \includegraphics[width=0.49\linewidth]{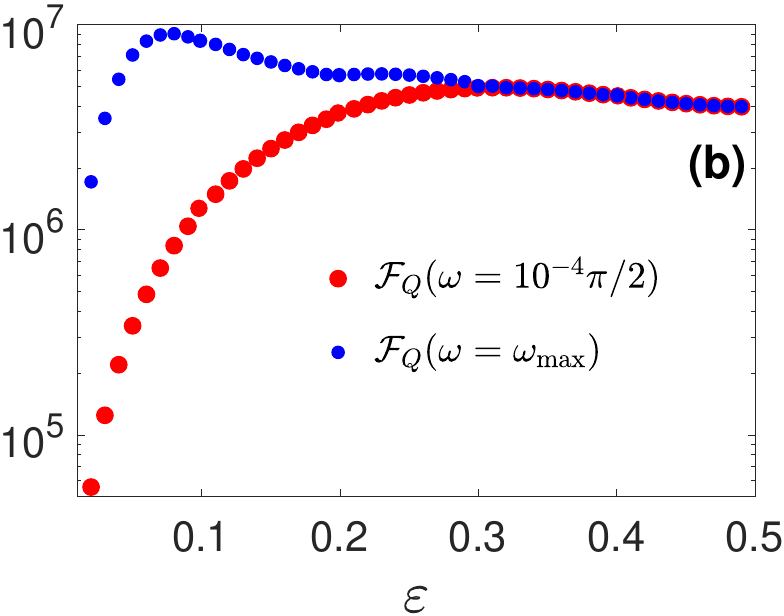}
    \caption{(a) The QFI as a function of the deviation $\omega$ for different values of the $\varepsilon$. (b) The extracted QFI as a function of \(\varepsilon\) deep in the DTC phase, \(\omega=10^{-4}\pi/2\), and at the finite-size threshold, \(\omega=\omega_{\max}\). In both panels, the results are obtained for a system of size $L=12$ with nonlinearity $\gamma{=}2$ after $n{=}10$ period cycle.}
    \label{fig:4}
\end{figure}
\section{The effect of imperfect pulse}\label{Sec.VI}
After discussing how nonlinearity improves sensing precision, this section focuses on investigating the impact of errors in the \(x\)-rotation pulse on the efficacy of the DTC probe.
Up to now, all the presented results are obtained for $\varepsilon{=}0.01$.
The QFI as a function of $\omega$ for different errors $\varepsilon$, in a system with nonlinear interaction $\gamma{=}2$, is illustrated in Fig.~\ref{fig:4}(a).  
Regardless of $\varepsilon$, one can observe the distinguishing behavior of the QFI within the DTC phase and the non-DTC region. 
While the qualitative behavior of the probe remains unaffected by \(\varepsilon\), the QFI value may increase as the pulse error $\varepsilon$ increases. This phenomenon can be interpreted as the influence of imperfect rotating pulses, which involve a larger sector of the Hilbert space in the system dynamics, thereby encoding more information about $\omega$ into the quantum state. 
Interestingly, the point of the transition from the DTC to the non-DTC region, i.e. $\omega_{\max}$, appears to be almost insensitive to the level of error.
To evaluate the performance across a broader error range, Fig.~\ref{fig:4}(b) displays $\mathcal{F}_{Q}$ as a function of $\varepsilon$ deep in the DTC phase, i.e. $\omega{=}10^{-4}\pi/2$, and at the finite-size threshold, i.e. $\omega{=}\omega_{\max}$. 
The enhancement in the DTC phase, characterized by strong localization, exhibits a significantly more pronounced effect compared to the finite-size threshold, where the dynamics already reflect the competition between localized DTC behavior and non-DTC response.
This finding contrasts with conventional sensors, where imperfections typically degrade sensing capabilities. 
The results in Fig.~\ref{fig:4} are obtained after $n{=}10$ cycling periods, for a chain of length $L{=}12$ with $\gamma{=}2$, initialized in $|\psi_{0}\rangle{=}|{\uparrow,\uparrow,\cdots,\uparrow}\rangle$.

\section{Experimental proposal}
The nonlinear DTC probe can be implemented naturally on a digital superconducting quantum simulator. 
The required Floquet operator in Eq.~(\ref{Eq.FloquetUnitary}) can be implemented using versatile gates. 
Since all terms in \(H_I^{(\gamma)}\) commute, the interaction part can be decomposed into nearest-neighbor controlled-phase gates,
\[
e^{-i(\pi/2-\omega)H_I^{(\gamma)}}
=
\prod_{j=1}^{L-1}
e^{-i\theta_j\sigma_j^z\sigma_{j+1}^z},
\qquad
\theta_j=(\pi/2-\omega)j^\gamma .
\]
Thus, one Floquet cycle consists of \(L-1\) tunable two-qubit \(ZZ\) gates with bond-dependent phases \(\theta_j\), followed by \(L\) single-qubit rotations \(e^{-i\Phi\sigma_j^x}\). 
This is similar to the digital strategy used in~\cite{Mi2022TimeCrystallineEigenstateOrder}. 
The only modification in the present protocol is the replacement of the random phase profile by the ordered profile \(\theta_j\propto j^\gamma\).
A typical experimental sequence is therefore straightforward. 
First, the qubits are initialized in a computational-basis product state, for example $|\psi_0\rangle=|\uparrow,\uparrow,\cdots,\uparrow\rangle$
Second, the Floquet unitary \(U_F(\omega,\varepsilon)\) is applied repeatedly for \(n\) cycles. 
Finally, the qubits are measured in the computational basis. 
From the resulting bit-string probabilities, one can extract the revival fidelity and the corresponding classical Fisher information. 
It can be shown that such projective measurements are sufficient to capture the quantum-enhanced sensitivity of the DTC probe~\cite{yousefjani2025discrete}. 
The same measurement strategy is applicable here because the nonlinear interaction changes only the bond-dependent phase angles, not the measurement basis.
The relevant experimental requirements are therefore: programmable nearest-neighbor \(ZZ\) phases, coherent single-qubit rotations, and repeated computational-basis readout. 
In superconducting circuits, these ingredients are standard. 
Superconducting qubits, therefore, provide the most direct near-term platform for realizing the nonlinear DTC sensing protocol.

\section{Conclusion}
We have shown that nonlinear Stark-type Ising interactions provide a direct route to enhancing the sensing capability of disorder-free DTC probes. Starting from a periodically kicked spin chain, we generalized the linear-gradient DTC sensor to interactions with spatial weights \(j^\gamma\) and analyzed how the nonlinearity exponent \(\gamma\) controls the scaling of the QFI. The analytical discussion establishes a rigorous seminorm ceiling for the attainable QFI and clarifies why the product-state protocol realizes a weaker, physically relevant scaling governed by local covariance in the DTC regime. Numerically, we find that the QFI keeps its quadratic long-time growth with the number of Floquet cycles while its system-size exponent increases approximately linearly with \(\gamma\). For \(\gamma=2,3,\) and \(4\), the extracted exponents show a clear enhancement over the linear-gradient case, confirming that nonlinear interactions improve the precision of estimating the detuning from the resonant condition. We also find that the finite-size stability window of the DTC phase shrinks algebraically with system size, and that stronger nonlinearities make the probe more sensitive to small deviations from resonance. Furthermore, we show that a controlled imperfection in the global spin-rotation pulse does not simply degrade the sensor; instead, for product-state initialization, it can help mix different interaction-energy sectors and increase the information encoded in the state. We also outlined a digital superconducting-qubit implementation, where the nonlinear Stark profile can be realized through programmable bond-dependent \(ZZ\) gates followed by global spin rotations. These results identify nonlinear DTC probes as experimentally accessible, product-state-based sensors whose precision can be enhanced by engineering the spatial profile of the interaction.

\section{Acknowledgment}
AB acknowledges support from the National Natural Science Foundation of China (grants No. W2541020, No. 12274059, No. 12574528 and No. 1251101297).


\begin{thebibliography}{75}%
\makeatletter
\providecommand \@ifxundefined [1]{%
 \@ifx{#1\undefined}
}%
\providecommand \@ifnum [1]{%
 \ifnum #1\expandafter \@firstoftwo
 \else \expandafter \@secondoftwo
 \fi
}%
\providecommand \@ifx [1]{%
 \ifx #1\expandafter \@firstoftwo
 \else \expandafter \@secondoftwo
 \fi
}%
\providecommand \natexlab [1]{#1}%
\providecommand \enquote  [1]{``#1''}%
\providecommand \bibnamefont  [1]{#1}%
\providecommand \bibfnamefont [1]{#1}%
\providecommand \citenamefont [1]{#1}%
\providecommand \href@noop [0]{\@secondoftwo}%
\providecommand \href [0]{\begingroup \@sanitize@url \@href}%
\providecommand \@href[1]{\@@startlink{#1}\@@href}%
\providecommand \@@href[1]{\endgroup#1\@@endlink}%
\providecommand \@sanitize@url [0]{\catcode `\\12\catcode `\$12\catcode `\&12\catcode `\#12\catcode `\^12\catcode `\_12\catcode `\%12\relax}%
\providecommand \@@startlink[1]{}%
\providecommand \@@endlink[0]{}%
\providecommand \url  [0]{\begingroup\@sanitize@url \@url }%
\providecommand \@url [1]{\endgroup\@href {#1}{\urlprefix }}%
\providecommand \urlprefix  [0]{URL }%
\providecommand \Eprint [0]{\href }%
\providecommand \doibase [0]{https://doi.org/}%
\providecommand \selectlanguage [0]{\@gobble}%
\providecommand \bibinfo  [0]{\@secondoftwo}%
\providecommand \bibfield  [0]{\@secondoftwo}%
\providecommand \translation [1]{[#1]}%
\providecommand \BibitemOpen [0]{}%
\providecommand \bibitemStop [0]{}%
\providecommand \bibitemNoStop [0]{.\EOS\space}%
\providecommand \EOS [0]{\spacefactor3000\relax}%
\providecommand \BibitemShut  [1]{\csname bibitem#1\endcsname}%
\let\auto@bib@innerbib\@empty
\bibitem [{\citenamefont {Wilczek}(2012)}]{wilczek2012quantum}%
  \BibitemOpen
  \bibfield  {author} {\bibinfo {author} {\bibfnamefont {F.}~\bibnamefont {Wilczek}},\ }\bibfield  {title} {\bibinfo {title} {Quantum time crystals},\ }\href {https://doi.org/10.1103/PhysRevLett.109.160401} {\bibfield  {journal} {\bibinfo  {journal} {Phys. Rev. Lett.}\ }\textbf {\bibinfo {volume} {109}},\ \bibinfo {pages} {160401} (\bibinfo {year} {2012})}\BibitemShut {NoStop}%
\bibitem [{\citenamefont {Bruno}(2013)}]{Bruno2013b}%
  \BibitemOpen
  \bibfield  {author} {\bibinfo {author} {\bibfnamefont {P.}~\bibnamefont {Bruno}},\ }\bibfield  {title} {\bibinfo {title} {Impossibility of spontaneously rotating time crystals: A no-go theorem},\ }\href {https://doi.org/10.1103/PhysRevLett.111.070402} {\bibfield  {journal} {\bibinfo  {journal} {Phys. Rev. Lett.}\ }\textbf {\bibinfo {volume} {111}},\ \bibinfo {pages} {070402} (\bibinfo {year} {2013})}\BibitemShut {NoStop}%
\bibitem [{\citenamefont {Watanabe}\ and\ \citenamefont {Oshikawa}(2015)}]{watanabe2015absence}%
  \BibitemOpen
  \bibfield  {author} {\bibinfo {author} {\bibfnamefont {H.}~\bibnamefont {Watanabe}}\ and\ \bibinfo {author} {\bibfnamefont {M.}~\bibnamefont {Oshikawa}},\ }\bibfield  {title} {\bibinfo {title} {Absence of quantum time crystals},\ }\href {https://doi.org/10.1103/PhysRevLett.114.251603} {\bibfield  {journal} {\bibinfo  {journal} {Phys. Rev. Lett.}\ }\textbf {\bibinfo {volume} {114}},\ \bibinfo {pages} {251603} (\bibinfo {year} {2015})}\BibitemShut {NoStop}%
\bibitem [{\citenamefont {Sacha}(2015)}]{Sacha2015}%
  \BibitemOpen
  \bibfield  {author} {\bibinfo {author} {\bibfnamefont {K.}~\bibnamefont {Sacha}},\ }\bibfield  {title} {\bibinfo {title} {Modeling spontaneous breaking of time-translation symmetry},\ }\href {https://doi.org/10.1103/PhysRevA.91.033617} {\bibfield  {journal} {\bibinfo  {journal} {Phys. Rev. A}\ }\textbf {\bibinfo {volume} {91}},\ \bibinfo {pages} {033617} (\bibinfo {year} {2015})}\BibitemShut {NoStop}%
\bibitem [{\citenamefont {Khemani}\ \emph {et~al.}(2016)\citenamefont {Khemani}, \citenamefont {Lazarides}, \citenamefont {Moessner},\ and\ \citenamefont {Sondhi}}]{khemani2016phase}%
  \BibitemOpen
  \bibfield  {author} {\bibinfo {author} {\bibfnamefont {V.}~\bibnamefont {Khemani}}, \bibinfo {author} {\bibfnamefont {A.}~\bibnamefont {Lazarides}}, \bibinfo {author} {\bibfnamefont {R.}~\bibnamefont {Moessner}},\ and\ \bibinfo {author} {\bibfnamefont {S.~L.}\ \bibnamefont {Sondhi}},\ }\bibfield  {title} {\bibinfo {title} {Phase structure of driven quantum systems},\ }\href {https://doi.org/10.1103/PhysRevLett.116.250401} {\bibfield  {journal} {\bibinfo  {journal} {Phys. Rev. Lett.}\ }\textbf {\bibinfo {volume} {116}},\ \bibinfo {pages} {250401} (\bibinfo {year} {2016})}\BibitemShut {NoStop}%
\bibitem [{\citenamefont {Else}\ \emph {et~al.}(2016)\citenamefont {Else}, \citenamefont {Bauer},\ and\ \citenamefont {Nayak}}]{else2016floquet}%
  \BibitemOpen
  \bibfield  {author} {\bibinfo {author} {\bibfnamefont {D.~V.}\ \bibnamefont {Else}}, \bibinfo {author} {\bibfnamefont {B.}~\bibnamefont {Bauer}},\ and\ \bibinfo {author} {\bibfnamefont {C.}~\bibnamefont {Nayak}},\ }\bibfield  {title} {\bibinfo {title} {Floquet time crystals},\ }\href {https://doi.org/10.1103/PhysRevLett.117.090402} {\bibfield  {journal} {\bibinfo  {journal} {Phys. Rev. Lett.}\ }\textbf {\bibinfo {volume} {117}},\ \bibinfo {pages} {090402} (\bibinfo {year} {2016})}\BibitemShut {NoStop}%
\bibitem [{\citenamefont {Yao}\ \emph {et~al.}(2017)\citenamefont {Yao}, \citenamefont {Potter}, \citenamefont {Potirniche},\ and\ \citenamefont {Vishwanath}}]{yao2017discrete}%
  \BibitemOpen
  \bibfield  {author} {\bibinfo {author} {\bibfnamefont {N.~Y.}\ \bibnamefont {Yao}}, \bibinfo {author} {\bibfnamefont {A.~C.}\ \bibnamefont {Potter}}, \bibinfo {author} {\bibfnamefont {I.-D.}\ \bibnamefont {Potirniche}},\ and\ \bibinfo {author} {\bibfnamefont {A.}~\bibnamefont {Vishwanath}},\ }\bibfield  {title} {\bibinfo {title} {Discrete time crystals: Rigidity, criticality, and realizations},\ }\href {https://doi.org/10.1103/PhysRevLett.118.030401} {\bibfield  {journal} {\bibinfo  {journal} {Phys. Rev. Lett.}\ }\textbf {\bibinfo {volume} {118}},\ \bibinfo {pages} {030401} (\bibinfo {year} {2017})}\BibitemShut {NoStop}%
\bibitem [{\citenamefont {Sacha}\ and\ \citenamefont {Zakrzewski}(2018)}]{sacha2017time}%
  \BibitemOpen
  \bibfield  {author} {\bibinfo {author} {\bibfnamefont {K.}~\bibnamefont {Sacha}}\ and\ \bibinfo {author} {\bibfnamefont {J.}~\bibnamefont {Zakrzewski}},\ }\bibfield  {title} {\bibinfo {title} {Time crystals: A review},\ }\href {https://doi.org/10.1088/1361-6633/aa8b38} {\bibfield  {journal} {\bibinfo  {journal} {Rep. Prog. Phys.}\ }\textbf {\bibinfo {volume} {81}},\ \bibinfo {pages} {016401} (\bibinfo {year} {2018})}\BibitemShut {NoStop}%
\bibitem [{\citenamefont {Zhang}\ \emph {et~al.}(2017)\citenamefont {Zhang}, \citenamefont {Hess}, \citenamefont {Kyprianidis}, \citenamefont {Becker}, \citenamefont {Lee}, \citenamefont {Smith}, \citenamefont {Pagano}, \citenamefont {Potirniche}, \citenamefont {Potter}, \citenamefont {Vishwanath}, \citenamefont {Yao},\ and\ \citenamefont {Monroe}}]{zhang2017observation}%
  \BibitemOpen
  \bibfield  {author} {\bibinfo {author} {\bibfnamefont {J.}~\bibnamefont {Zhang}}, \bibinfo {author} {\bibfnamefont {P.~W.}\ \bibnamefont {Hess}}, \bibinfo {author} {\bibfnamefont {A.}~\bibnamefont {Kyprianidis}}, \bibinfo {author} {\bibfnamefont {P.}~\bibnamefont {Becker}}, \bibinfo {author} {\bibfnamefont {A.}~\bibnamefont {Lee}}, \bibinfo {author} {\bibfnamefont {J.}~\bibnamefont {Smith}}, \bibinfo {author} {\bibfnamefont {G.}~\bibnamefont {Pagano}}, \bibinfo {author} {\bibfnamefont {I.-D.}\ \bibnamefont {Potirniche}}, \bibinfo {author} {\bibfnamefont {A.~C.}\ \bibnamefont {Potter}}, \bibinfo {author} {\bibfnamefont {A.}~\bibnamefont {Vishwanath}}, \bibinfo {author} {\bibfnamefont {N.~Y.}\ \bibnamefont {Yao}},\ and\ \bibinfo {author} {\bibfnamefont {C.}~\bibnamefont {Monroe}},\ }\bibfield  {title} {\bibinfo {title} {Observation of a discrete time crystal},\ }\href {https://doi.org/10.1038/nature21413} {\bibfield  {journal} {\bibinfo  {journal} {Nature}\ }\textbf {\bibinfo {volume} {543}},\ \bibinfo
  {pages} {217} (\bibinfo {year} {2017})}\BibitemShut {NoStop}%
\bibitem [{\citenamefont {Choi}\ \emph {et~al.}(2017)\citenamefont {Choi}, \citenamefont {Choi}, \citenamefont {Landig}, \citenamefont {Kucsko}, \citenamefont {Zhou}, \citenamefont {Isoya}, \citenamefont {Jelezko}, \citenamefont {Onoda}, \citenamefont {Sumiya}, \citenamefont {Khemani}, \citenamefont {von Keyserlingk}, \citenamefont {Yao}, \citenamefont {Demler},\ and\ \citenamefont {Lukin}}]{choi2017observation}%
  \BibitemOpen
  \bibfield  {author} {\bibinfo {author} {\bibfnamefont {S.}~\bibnamefont {Choi}}, \bibinfo {author} {\bibfnamefont {J.}~\bibnamefont {Choi}}, \bibinfo {author} {\bibfnamefont {R.}~\bibnamefont {Landig}}, \bibinfo {author} {\bibfnamefont {G.}~\bibnamefont {Kucsko}}, \bibinfo {author} {\bibfnamefont {H.}~\bibnamefont {Zhou}}, \bibinfo {author} {\bibfnamefont {J.}~\bibnamefont {Isoya}}, \bibinfo {author} {\bibfnamefont {F.}~\bibnamefont {Jelezko}}, \bibinfo {author} {\bibfnamefont {S.}~\bibnamefont {Onoda}}, \bibinfo {author} {\bibfnamefont {H.}~\bibnamefont {Sumiya}}, \bibinfo {author} {\bibfnamefont {V.}~\bibnamefont {Khemani}}, \bibinfo {author} {\bibfnamefont {C.}~\bibnamefont {von Keyserlingk}}, \bibinfo {author} {\bibfnamefont {N.~Y.}\ \bibnamefont {Yao}}, \bibinfo {author} {\bibfnamefont {E.}~\bibnamefont {Demler}},\ and\ \bibinfo {author} {\bibfnamefont {M.~D.}\ \bibnamefont {Lukin}},\ }\bibfield  {title} {\bibinfo {title} {Observation of discrete time-crystalline order in a disordered dipolar many-body
  system},\ }\href {https://doi.org/10.1038/nature21426} {\bibfield  {journal} {\bibinfo  {journal} {Nature}\ }\textbf {\bibinfo {volume} {543}},\ \bibinfo {pages} {221} (\bibinfo {year} {2017})}\BibitemShut {NoStop}%
\bibitem [{\citenamefont {Rovny}\ \emph {et~al.}(2018)\citenamefont {Rovny}, \citenamefont {Blum},\ and\ \citenamefont {Barrett}}]{rovny2018observation}%
  \BibitemOpen
  \bibfield  {author} {\bibinfo {author} {\bibfnamefont {J.}~\bibnamefont {Rovny}}, \bibinfo {author} {\bibfnamefont {R.~L.}\ \bibnamefont {Blum}},\ and\ \bibinfo {author} {\bibfnamefont {S.~E.}\ \bibnamefont {Barrett}},\ }\bibfield  {title} {\bibinfo {title} {Observation of discrete-time-crystal signatures in an ordered dipolar many-body system},\ }\href {https://doi.org/10.1103/PhysRevLett.120.180603} {\bibfield  {journal} {\bibinfo  {journal} {Phys. Rev. Lett.}\ }\textbf {\bibinfo {volume} {120}},\ \bibinfo {pages} {180603} (\bibinfo {year} {2018})}\BibitemShut {NoStop}%
\bibitem [{\citenamefont {Liu}\ \emph {et~al.}(2024)\citenamefont {Liu}, \citenamefont {Zhang}, \citenamefont {Wang}, \citenamefont {Ma}, \citenamefont {Han}, \citenamefont {Zhang}, \citenamefont {Zhang}, \citenamefont {Shao}, \citenamefont {Li}, \citenamefont {Chen}, \citenamefont {Shi},\ and\ \citenamefont {Ding}}]{Liu2024RydbergDTC}%
  \BibitemOpen
  \bibfield  {author} {\bibinfo {author} {\bibfnamefont {B.}~\bibnamefont {Liu}}, \bibinfo {author} {\bibfnamefont {L.-H.}\ \bibnamefont {Zhang}}, \bibinfo {author} {\bibfnamefont {Q.-F.}\ \bibnamefont {Wang}}, \bibinfo {author} {\bibfnamefont {Y.}~\bibnamefont {Ma}}, \bibinfo {author} {\bibfnamefont {T.-Y.}\ \bibnamefont {Han}}, \bibinfo {author} {\bibfnamefont {J.}~\bibnamefont {Zhang}}, \bibinfo {author} {\bibfnamefont {Z.-Y.}\ \bibnamefont {Zhang}}, \bibinfo {author} {\bibfnamefont {S.-Y.}\ \bibnamefont {Shao}}, \bibinfo {author} {\bibfnamefont {Q.}~\bibnamefont {Li}}, \bibinfo {author} {\bibfnamefont {H.-C.}\ \bibnamefont {Chen}}, \bibinfo {author} {\bibfnamefont {B.-S.}\ \bibnamefont {Shi}},\ and\ \bibinfo {author} {\bibfnamefont {D.-S.}\ \bibnamefont {Ding}},\ }\bibfield  {title} {\bibinfo {title} {Higher-order and fractional discrete time crystals in floquet-driven rydberg atoms},\ }\href {https://doi.org/10.1038/s41467-024-53712-5} {\bibfield  {journal} {\bibinfo  {journal} {Nat. Commun.}\ }\textbf
  {\bibinfo {volume} {15}},\ \bibinfo {pages} {9730} (\bibinfo {year} {2024})}\BibitemShut {NoStop}%
\bibitem [{\citenamefont {Liu}\ \emph {et~al.}(2025)\citenamefont {Liu}, \citenamefont {Zhang}, \citenamefont {Ma}, \citenamefont {Wang}, \citenamefont {Han}, \citenamefont {Zhang}, \citenamefont {Zhang}, \citenamefont {Shao}, \citenamefont {Li}, \citenamefont {Chen}, \citenamefont {Guo}, \citenamefont {Ding},\ and\ \citenamefont {Shi}}]{Liu2025BifurcationTimeCrystalsRydberg}%
  \BibitemOpen
  \bibfield  {author} {\bibinfo {author} {\bibfnamefont {B.}~\bibnamefont {Liu}}, \bibinfo {author} {\bibfnamefont {L.-H.}\ \bibnamefont {Zhang}}, \bibinfo {author} {\bibfnamefont {Y.}~\bibnamefont {Ma}}, \bibinfo {author} {\bibfnamefont {Q.-F.}\ \bibnamefont {Wang}}, \bibinfo {author} {\bibfnamefont {T.-Y.}\ \bibnamefont {Han}}, \bibinfo {author} {\bibfnamefont {J.}~\bibnamefont {Zhang}}, \bibinfo {author} {\bibfnamefont {Z.-Y.}\ \bibnamefont {Zhang}}, \bibinfo {author} {\bibfnamefont {S.-Y.}\ \bibnamefont {Shao}}, \bibinfo {author} {\bibfnamefont {Q.}~\bibnamefont {Li}}, \bibinfo {author} {\bibfnamefont {H.-C.}\ \bibnamefont {Chen}}, \bibinfo {author} {\bibfnamefont {G.-C.}\ \bibnamefont {Guo}}, \bibinfo {author} {\bibfnamefont {D.-S.}\ \bibnamefont {Ding}},\ and\ \bibinfo {author} {\bibfnamefont {B.-S.}\ \bibnamefont {Shi}},\ }\bibfield  {title} {\bibinfo {title} {Bifurcation of time crystals in driven and dissipative rydberg atomic gas},\ }\href {https://doi.org/10.1038/s41467-025-56712-1} {\bibfield
  {journal} {\bibinfo  {journal} {Nat. Commun.}\ }\textbf {\bibinfo {volume} {16}},\ \bibinfo {pages} {1419} (\bibinfo {year} {2025})}\BibitemShut {NoStop}%
\bibitem [{\citenamefont {Liu}\ \emph {et~al.}(2026)\citenamefont {Liu}, \citenamefont {Chen}, \citenamefont {Ma}, \citenamefont {Wang}, \citenamefont {Han}, \citenamefont {Tian}, \citenamefont {Qian}, \citenamefont {Guo}, \citenamefont {Zhang}, \citenamefont {Wei}, \citenamefont {Bayat}, \citenamefont {Ding},\ and\ \citenamefont {Shi}}]{Liu2026EnhancedMultiparameterMetrology}%
  \BibitemOpen
  \bibfield  {author} {\bibinfo {author} {\bibfnamefont {B.}~\bibnamefont {Liu}}, \bibinfo {author} {\bibfnamefont {J.-R.}\ \bibnamefont {Chen}}, \bibinfo {author} {\bibfnamefont {Y.}~\bibnamefont {Ma}}, \bibinfo {author} {\bibfnamefont {Q.-F.}\ \bibnamefont {Wang}}, \bibinfo {author} {\bibfnamefont {T.-Y.}\ \bibnamefont {Han}}, \bibinfo {author} {\bibfnamefont {H.}~\bibnamefont {Tian}}, \bibinfo {author} {\bibfnamefont {Y.-H.}\ \bibnamefont {Qian}}, \bibinfo {author} {\bibfnamefont {G.-C.}\ \bibnamefont {Guo}}, \bibinfo {author} {\bibfnamefont {L.-H.}\ \bibnamefont {Zhang}}, \bibinfo {author} {\bibfnamefont {B.-B.}\ \bibnamefont {Wei}}, \bibinfo {author} {\bibfnamefont {A.}~\bibnamefont {Bayat}}, \bibinfo {author} {\bibfnamefont {D.-S.}\ \bibnamefont {Ding}},\ and\ \bibinfo {author} {\bibfnamefont {B.-S.}\ \bibnamefont {Shi}},\ }\href {https://doi.org/10.48550/arXiv.2601.10347} {\bibinfo {title} {Enhanced multi-parameter metrology in dissipative rydberg atom time crystals}} (\bibinfo {year} {2026}),\ \Eprint
  {https://arxiv.org/abs/2601.10347} {arXiv:2601.10347} \BibitemShut {NoStop}%
\bibitem [{\citenamefont {Zhu}\ \emph {et~al.}(2025)\citenamefont {Zhu}, \citenamefont {Zhang}, \citenamefont {Wang}, \citenamefont {Ma}, \citenamefont {Han}, \citenamefont {Yu}, \citenamefont {Fang}, \citenamefont {Shao}, \citenamefont {Li}, \citenamefont {Wang}, \citenamefont {Zhang}, \citenamefont {Chen}, \citenamefont {Liu}, \citenamefont {Nan}, \citenamefont {Yin}, \citenamefont {Zhang}, \citenamefont {Guo}, \citenamefont {Liu}, \citenamefont {Ding},\ and\ \citenamefont {Shi}}]{Zhu2025DiscreteTimeQuasicrystalRydberg}%
  \BibitemOpen
  \bibfield  {author} {\bibinfo {author} {\bibfnamefont {D.-Y.}\ \bibnamefont {Zhu}}, \bibinfo {author} {\bibfnamefont {Z.-Y.}\ \bibnamefont {Zhang}}, \bibinfo {author} {\bibfnamefont {Q.-F.}\ \bibnamefont {Wang}}, \bibinfo {author} {\bibfnamefont {Y.}~\bibnamefont {Ma}}, \bibinfo {author} {\bibfnamefont {T.-Y.}\ \bibnamefont {Han}}, \bibinfo {author} {\bibfnamefont {C.}~\bibnamefont {Yu}}, \bibinfo {author} {\bibfnamefont {Q.-Q.}\ \bibnamefont {Fang}}, \bibinfo {author} {\bibfnamefont {S.-Y.}\ \bibnamefont {Shao}}, \bibinfo {author} {\bibfnamefont {Q.}~\bibnamefont {Li}}, \bibinfo {author} {\bibfnamefont {Y.-J.}\ \bibnamefont {Wang}}, \bibinfo {author} {\bibfnamefont {J.}~\bibnamefont {Zhang}}, \bibinfo {author} {\bibfnamefont {H.-C.}\ \bibnamefont {Chen}}, \bibinfo {author} {\bibfnamefont {X.}~\bibnamefont {Liu}}, \bibinfo {author} {\bibfnamefont {J.-D.}\ \bibnamefont {Nan}}, \bibinfo {author} {\bibfnamefont {Y.-M.}\ \bibnamefont {Yin}}, \bibinfo {author} {\bibfnamefont {L.-H.}\ \bibnamefont {Zhang}},
  \bibinfo {author} {\bibfnamefont {G.-C.}\ \bibnamefont {Guo}}, \bibinfo {author} {\bibfnamefont {B.}~\bibnamefont {Liu}}, \bibinfo {author} {\bibfnamefont {D.-S.}\ \bibnamefont {Ding}},\ and\ \bibinfo {author} {\bibfnamefont {B.-S.}\ \bibnamefont {Shi}},\ }\href {https://doi.org/10.48550/arXiv.2509.21248} {\bibinfo {title} {Observation of discrete time quasicrystal in rydberg atomic gases}} (\bibinfo {year} {2025}),\ \Eprint {https://arxiv.org/abs/2509.21248} {arXiv:2509.21248} \BibitemShut {NoStop}%
\bibitem [{\citenamefont {Alet}\ and\ \citenamefont {Laflorencie}(2018)}]{alet2018many}%
  \BibitemOpen
  \bibfield  {author} {\bibinfo {author} {\bibfnamefont {F.}~\bibnamefont {Alet}}\ and\ \bibinfo {author} {\bibfnamefont {N.}~\bibnamefont {Laflorencie}},\ }\bibfield  {title} {\bibinfo {title} {Many-body localization: An introduction and selected topics},\ }\href {https://doi.org/10.1016/j.crhy.2018.03.003} {\bibfield  {journal} {\bibinfo  {journal} {Comptes Rendus Physique}\ }\textbf {\bibinfo {volume} {19}},\ \bibinfo {pages} {498} (\bibinfo {year} {2018})}\BibitemShut {NoStop}%
\bibitem [{\citenamefont {Schulz}\ \emph {et~al.}(2019)\citenamefont {Schulz}, \citenamefont {Hooley}, \citenamefont {Moessner},\ and\ \citenamefont {Pollmann}}]{schulz2019stark}%
  \BibitemOpen
  \bibfield  {author} {\bibinfo {author} {\bibfnamefont {M.}~\bibnamefont {Schulz}}, \bibinfo {author} {\bibfnamefont {C.}~\bibnamefont {Hooley}}, \bibinfo {author} {\bibfnamefont {R.}~\bibnamefont {Moessner}},\ and\ \bibinfo {author} {\bibfnamefont {F.}~\bibnamefont {Pollmann}},\ }\bibfield  {title} {\bibinfo {title} {{Stark} many-body localization},\ }\href {https://doi.org/10.1103/PhysRevLett.122.040606} {\bibfield  {journal} {\bibinfo  {journal} {Phys. Rev. Lett.}\ }\textbf {\bibinfo {volume} {122}},\ \bibinfo {pages} {040606} (\bibinfo {year} {2019})}\BibitemShut {NoStop}%
\bibitem [{\citenamefont {Morong}\ \emph {et~al.}(2021)\citenamefont {Morong}, \citenamefont {Liu}, \citenamefont {Becker}, \citenamefont {Collins}, \citenamefont {Feng}, \citenamefont {Kyprianidis}, \citenamefont {Pagano}, \citenamefont {You}, \citenamefont {Gorshkov},\ and\ \citenamefont {Monroe}}]{morong2021observation}%
  \BibitemOpen
  \bibfield  {author} {\bibinfo {author} {\bibfnamefont {W.}~\bibnamefont {Morong}}, \bibinfo {author} {\bibfnamefont {F.}~\bibnamefont {Liu}}, \bibinfo {author} {\bibfnamefont {P.}~\bibnamefont {Becker}}, \bibinfo {author} {\bibfnamefont {K.~S.}\ \bibnamefont {Collins}}, \bibinfo {author} {\bibfnamefont {L.}~\bibnamefont {Feng}}, \bibinfo {author} {\bibfnamefont {A.}~\bibnamefont {Kyprianidis}}, \bibinfo {author} {\bibfnamefont {G.}~\bibnamefont {Pagano}}, \bibinfo {author} {\bibfnamefont {T.}~\bibnamefont {You}}, \bibinfo {author} {\bibfnamefont {A.~V.}\ \bibnamefont {Gorshkov}},\ and\ \bibinfo {author} {\bibfnamefont {C.}~\bibnamefont {Monroe}},\ }\bibfield  {title} {\bibinfo {title} {Observation of {Stark} many-body localization without disorder},\ }\href {https://doi.org/10.1038/s41586-021-03988-0} {\bibfield  {journal} {\bibinfo  {journal} {Nature}\ }\textbf {\bibinfo {volume} {599}},\ \bibinfo {pages} {393} (\bibinfo {year} {2021})}\BibitemShut {NoStop}%
\bibitem [{\citenamefont {Yousefjani}\ and\ \citenamefont {Bayat}(2023)}]{Rozha1}%
  \BibitemOpen
  \bibfield  {author} {\bibinfo {author} {\bibfnamefont {R.}~\bibnamefont {Yousefjani}}\ and\ \bibinfo {author} {\bibfnamefont {A.}~\bibnamefont {Bayat}},\ }\bibfield  {title} {\bibinfo {title} {Mobility edge in long-range interacting many-body localized systems},\ }\href {https://doi.org/10.1103/PhysRevB.107.045108} {\bibfield  {journal} {\bibinfo  {journal} {Phys. Rev. B}\ }\textbf {\bibinfo {volume} {107}},\ \bibinfo {pages} {045108} (\bibinfo {year} {2023})}\BibitemShut {NoStop}%
\bibitem [{\citenamefont {Yousefjani}\ \emph {et~al.}(2023{\natexlab{a}})\citenamefont {Yousefjani}, \citenamefont {Bose},\ and\ \citenamefont {Bayat}}]{Rozha2}%
  \BibitemOpen
  \bibfield  {author} {\bibinfo {author} {\bibfnamefont {R.}~\bibnamefont {Yousefjani}}, \bibinfo {author} {\bibfnamefont {S.}~\bibnamefont {Bose}},\ and\ \bibinfo {author} {\bibfnamefont {A.}~\bibnamefont {Bayat}},\ }\bibfield  {title} {\bibinfo {title} {Floquet-induced localization in long-range many-body systems},\ }\href {https://doi.org/10.1103/PhysRevResearch.5.013094} {\bibfield  {journal} {\bibinfo  {journal} {Phys. Rev. Res.}\ }\textbf {\bibinfo {volume} {5}},\ \bibinfo {pages} {013094} (\bibinfo {year} {2023}{\natexlab{a}})}\BibitemShut {NoStop}%
\bibitem [{\citenamefont {Kshetrimayum}\ \emph {et~al.}(2020)\citenamefont {Kshetrimayum}, \citenamefont {Eisert},\ and\ \citenamefont {Kennes}}]{kshetrimayum2020stark}%
  \BibitemOpen
  \bibfield  {author} {\bibinfo {author} {\bibfnamefont {A.}~\bibnamefont {Kshetrimayum}}, \bibinfo {author} {\bibfnamefont {J.}~\bibnamefont {Eisert}},\ and\ \bibinfo {author} {\bibfnamefont {D.}~\bibnamefont {Kennes}},\ }\bibfield  {title} {\bibinfo {title} {Stark time crystals: Symmetry breaking in space and time},\ }\href {https://doi.org/10.1103/PhysRevB.102.195116} {\bibfield  {journal} {\bibinfo  {journal} {Phys. Rev. B}\ }\textbf {\bibinfo {volume} {102}},\ \bibinfo {pages} {195116} (\bibinfo {year} {2020})}\BibitemShut {NoStop}%
\bibitem [{\citenamefont {Liu}\ \emph {et~al.}(2023)\citenamefont {Liu}, \citenamefont {Zhang}, \citenamefont {Hsieh}, \citenamefont {Zhang},\ and\ \citenamefont {Yao}}]{liu2023discrete}%
  \BibitemOpen
  \bibfield  {author} {\bibinfo {author} {\bibfnamefont {S.}~\bibnamefont {Liu}}, \bibinfo {author} {\bibfnamefont {S.-X.}\ \bibnamefont {Zhang}}, \bibinfo {author} {\bibfnamefont {C.-Y.}\ \bibnamefont {Hsieh}}, \bibinfo {author} {\bibfnamefont {S.}~\bibnamefont {Zhang}},\ and\ \bibinfo {author} {\bibfnamefont {H.}~\bibnamefont {Yao}},\ }\bibfield  {title} {\bibinfo {title} {Discrete time crystal enabled by stark many-body localization},\ }\href {https://doi.org/10.1103/PhysRevLett.130.120403} {\bibfield  {journal} {\bibinfo  {journal} {Phys. Rev. Lett.}\ }\textbf {\bibinfo {volume} {130}},\ \bibinfo {pages} {120403} (\bibinfo {year} {2023})}\BibitemShut {NoStop}%
\bibitem [{\citenamefont {Sajid}\ \emph {et~al.}(2025)\citenamefont {Sajid}, \citenamefont {Yousefjani},\ and\ \citenamefont {Bayat}}]{sajid2025thermal}%
  \BibitemOpen
  \bibfield  {author} {\bibinfo {author} {\bibfnamefont {M.}~\bibnamefont {Sajid}}, \bibinfo {author} {\bibfnamefont {R.}~\bibnamefont {Yousefjani}},\ and\ \bibinfo {author} {\bibfnamefont {A.}~\bibnamefont {Bayat}},\ }\bibfield  {title} {\bibinfo {title} {Thermal avalanches in isolated many-body localized systems},\ }\href {https://doi.org/10.1103/fvgj-byp7} {\bibfield  {journal} {\bibinfo  {journal} {Phys. Rev. B}\ }\textbf {\bibinfo {volume} {112}},\ \bibinfo {pages} {155140} (\bibinfo {year} {2025})}\BibitemShut {NoStop}%
\bibitem [{\citenamefont {Raghunandan}\ \emph {et~al.}(2018)\citenamefont {Raghunandan}, \citenamefont {Wrachtrup},\ and\ \citenamefont {Weimer}}]{raghunandan2018high}%
  \BibitemOpen
  \bibfield  {author} {\bibinfo {author} {\bibfnamefont {M.}~\bibnamefont {Raghunandan}}, \bibinfo {author} {\bibfnamefont {J.}~\bibnamefont {Wrachtrup}},\ and\ \bibinfo {author} {\bibfnamefont {H.}~\bibnamefont {Weimer}},\ }\bibfield  {title} {\bibinfo {title} {High-density quantum sensing with dissipative first order transitions},\ }\href {https://doi.org/10.1103/PhysRevLett.120.150501} {\bibfield  {journal} {\bibinfo  {journal} {Phys. Rev. Lett.}\ }\textbf {\bibinfo {volume} {120}},\ \bibinfo {pages} {150501} (\bibinfo {year} {2018})}\BibitemShut {NoStop}%
\bibitem [{\citenamefont {Heugel}\ \emph {et~al.}(2019)\citenamefont {Heugel}, \citenamefont {Biondi}, \citenamefont {Zilberberg},\ and\ \citenamefont {Chitra}}]{heugel2019quantum}%
  \BibitemOpen
  \bibfield  {author} {\bibinfo {author} {\bibfnamefont {T.~L.}\ \bibnamefont {Heugel}}, \bibinfo {author} {\bibfnamefont {M.}~\bibnamefont {Biondi}}, \bibinfo {author} {\bibfnamefont {O.}~\bibnamefont {Zilberberg}},\ and\ \bibinfo {author} {\bibfnamefont {R.}~\bibnamefont {Chitra}},\ }\bibfield  {title} {\bibinfo {title} {Quantum transducer using a parametric driven-dissipative phase transition},\ }\href {https://doi.org/10.1103/PhysRevLett.123.173601} {\bibfield  {journal} {\bibinfo  {journal} {Phys. Rev. Lett.}\ }\textbf {\bibinfo {volume} {123}},\ \bibinfo {pages} {173601} (\bibinfo {year} {2019})}\BibitemShut {NoStop}%
\bibitem [{\citenamefont {Yang}\ and\ \citenamefont {Jacob}(2019)}]{yang2019engineering}%
  \BibitemOpen
  \bibfield  {author} {\bibinfo {author} {\bibfnamefont {L.-P.}\ \bibnamefont {Yang}}\ and\ \bibinfo {author} {\bibfnamefont {Z.}~\bibnamefont {Jacob}},\ }\bibfield  {title} {\bibinfo {title} {Engineering first-order quantum phase transitions for weak signal detection},\ }\href {https://doi.org/10.1063/1.5121558} {\bibfield  {journal} {\bibinfo  {journal} {J. Appl. Phys.}\ }\textbf {\bibinfo {volume} {126}},\ \bibinfo {pages} {174502} (\bibinfo {year} {2019})}\BibitemShut {NoStop}%
\bibitem [{\citenamefont {Zanardi}\ \emph {et~al.}(2008)\citenamefont {Zanardi}, \citenamefont {Paris},\ and\ \citenamefont {Venuti}}]{zanardi2008quantum}%
  \BibitemOpen
  \bibfield  {author} {\bibinfo {author} {\bibfnamefont {P.}~\bibnamefont {Zanardi}}, \bibinfo {author} {\bibfnamefont {M.~G.}\ \bibnamefont {Paris}},\ and\ \bibinfo {author} {\bibfnamefont {L.~C.}\ \bibnamefont {Venuti}},\ }\bibfield  {title} {\bibinfo {title} {Quantum criticality as a resource for quantum estimation},\ }\href {https://doi.org/10.1103/PhysRevA.78.042105} {\bibfield  {journal} {\bibinfo  {journal} {Phys. Rev. A}\ }\textbf {\bibinfo {volume} {78}},\ \bibinfo {pages} {042105} (\bibinfo {year} {2008})}\BibitemShut {NoStop}%
\bibitem [{\citenamefont {Invernizzi}\ \emph {et~al.}(2008)\citenamefont {Invernizzi}, \citenamefont {Korbman}, \citenamefont {Venuti},\ and\ \citenamefont {Paris}}]{invernizzi2008optimal}%
  \BibitemOpen
  \bibfield  {author} {\bibinfo {author} {\bibfnamefont {C.}~\bibnamefont {Invernizzi}}, \bibinfo {author} {\bibfnamefont {M.}~\bibnamefont {Korbman}}, \bibinfo {author} {\bibfnamefont {L.~C.}\ \bibnamefont {Venuti}},\ and\ \bibinfo {author} {\bibfnamefont {M.~G.}\ \bibnamefont {Paris}},\ }\bibfield  {title} {\bibinfo {title} {Optimal quantum estimation in spin systems at criticality},\ }\href {https://doi.org/10.1103/PhysRevA.78.042106} {\bibfield  {journal} {\bibinfo  {journal} {Phys. Rev. A}\ }\textbf {\bibinfo {volume} {78}},\ \bibinfo {pages} {042106} (\bibinfo {year} {2008})}\BibitemShut {NoStop}%
\bibitem [{\citenamefont {Chu}\ \emph {et~al.}(2021)\citenamefont {Chu}, \citenamefont {Zhang}, \citenamefont {Yu},\ and\ \citenamefont {Cai}}]{chu2021dynamic}%
  \BibitemOpen
  \bibfield  {author} {\bibinfo {author} {\bibfnamefont {Y.}~\bibnamefont {Chu}}, \bibinfo {author} {\bibfnamefont {S.}~\bibnamefont {Zhang}}, \bibinfo {author} {\bibfnamefont {B.}~\bibnamefont {Yu}},\ and\ \bibinfo {author} {\bibfnamefont {J.}~\bibnamefont {Cai}},\ }\bibfield  {title} {\bibinfo {title} {Dynamic framework for criticality-enhanced quantum sensing},\ }\href {https://doi.org/10.1103/PhysRevLett.126.010502} {\bibfield  {journal} {\bibinfo  {journal} {Phys. Rev. Lett.}\ }\textbf {\bibinfo {volume} {126}},\ \bibinfo {pages} {010502} (\bibinfo {year} {2021})}\BibitemShut {NoStop}%
\bibitem [{\citenamefont {Montenegro}\ \emph {et~al.}(2021)\citenamefont {Montenegro}, \citenamefont {Mishra},\ and\ \citenamefont {Bayat}}]{montenegro2021global}%
  \BibitemOpen
  \bibfield  {author} {\bibinfo {author} {\bibfnamefont {V.}~\bibnamefont {Montenegro}}, \bibinfo {author} {\bibfnamefont {U.}~\bibnamefont {Mishra}},\ and\ \bibinfo {author} {\bibfnamefont {A.}~\bibnamefont {Bayat}},\ }\bibfield  {title} {\bibinfo {title} {Global sensing and its impact for quantum many-body probes with criticality},\ }\href {https://doi.org/10.1103/PhysRevLett.126.200501} {\bibfield  {journal} {\bibinfo  {journal} {Phys. Rev. Lett.}\ }\textbf {\bibinfo {volume} {126}},\ \bibinfo {pages} {200501} (\bibinfo {year} {2021})}\BibitemShut {NoStop}%
\bibitem [{\citenamefont {Fern{\'a}ndez-Lorenzo}\ and\ \citenamefont {Porras}(2017)}]{fernandez2017quantum}%
  \BibitemOpen
  \bibfield  {author} {\bibinfo {author} {\bibfnamefont {S.}~\bibnamefont {Fern{\'a}ndez-Lorenzo}}\ and\ \bibinfo {author} {\bibfnamefont {D.}~\bibnamefont {Porras}},\ }\bibfield  {title} {\bibinfo {title} {Quantum sensing close to a dissipative phase transition: Symmetry breaking and criticality as metrological resources},\ }\href {https://doi.org/10.1103/PhysRevA.96.013817} {\bibfield  {journal} {\bibinfo  {journal} {Phys. Rev. A}\ }\textbf {\bibinfo {volume} {96}},\ \bibinfo {pages} {013817} (\bibinfo {year} {2017})}\BibitemShut {NoStop}%
\bibitem [{\citenamefont {Ilias}\ \emph {et~al.}(2022)\citenamefont {Ilias}, \citenamefont {Yang}, \citenamefont {Huelga},\ and\ \citenamefont {Plenio}}]{ilias2022criticality}%
  \BibitemOpen
  \bibfield  {author} {\bibinfo {author} {\bibfnamefont {T.}~\bibnamefont {Ilias}}, \bibinfo {author} {\bibfnamefont {D.}~\bibnamefont {Yang}}, \bibinfo {author} {\bibfnamefont {S.~F.}\ \bibnamefont {Huelga}},\ and\ \bibinfo {author} {\bibfnamefont {M.~B.}\ \bibnamefont {Plenio}},\ }\bibfield  {title} {\bibinfo {title} {Criticality-enhanced quantum sensing via continuous measurement},\ }\href {https://doi.org/10.1103/PRXQuantum.3.010354} {\bibfield  {journal} {\bibinfo  {journal} {PRX Quantum}\ }\textbf {\bibinfo {volume} {3}},\ \bibinfo {pages} {010354} (\bibinfo {year} {2022})}\BibitemShut {NoStop}%
\bibitem [{\citenamefont {Sarkar}\ \emph {et~al.}(2022)\citenamefont {Sarkar}, \citenamefont {Mukhopadhyay}, \citenamefont {Alase},\ and\ \citenamefont {Bayat}}]{sarkar2022free}%
  \BibitemOpen
  \bibfield  {author} {\bibinfo {author} {\bibfnamefont {S.}~\bibnamefont {Sarkar}}, \bibinfo {author} {\bibfnamefont {C.}~\bibnamefont {Mukhopadhyay}}, \bibinfo {author} {\bibfnamefont {A.}~\bibnamefont {Alase}},\ and\ \bibinfo {author} {\bibfnamefont {A.}~\bibnamefont {Bayat}},\ }\bibfield  {title} {\bibinfo {title} {Free-fermionic topological quantum sensors},\ }\href {https://doi.org/10.1103/PhysRevLett.129.090503} {\bibfield  {journal} {\bibinfo  {journal} {Phys. Rev. Lett.}\ }\textbf {\bibinfo {volume} {129}},\ \bibinfo {pages} {090503} (\bibinfo {year} {2022})}\BibitemShut {NoStop}%
\bibitem [{\citenamefont {Budich}\ and\ \citenamefont {Bergholtz}(2020)}]{budich2020non}%
  \BibitemOpen
  \bibfield  {author} {\bibinfo {author} {\bibfnamefont {J.~C.}\ \bibnamefont {Budich}}\ and\ \bibinfo {author} {\bibfnamefont {E.~J.}\ \bibnamefont {Bergholtz}},\ }\bibfield  {title} {\bibinfo {title} {Non-{H}ermitian topological sensors},\ }\href {https://doi.org/10.1103/PhysRevLett.125.180403} {\bibfield  {journal} {\bibinfo  {journal} {Phys. Rev. Lett.}\ }\textbf {\bibinfo {volume} {125}},\ \bibinfo {pages} {180403} (\bibinfo {year} {2020})}\BibitemShut {NoStop}%
\bibitem [{\citenamefont {Koch}\ and\ \citenamefont {Budich}(2022)}]{koch2022quantum}%
  \BibitemOpen
  \bibfield  {author} {\bibinfo {author} {\bibfnamefont {F.}~\bibnamefont {Koch}}\ and\ \bibinfo {author} {\bibfnamefont {J.~C.}\ \bibnamefont {Budich}},\ }\bibfield  {title} {\bibinfo {title} {Quantum non-{H}ermitian topological sensors},\ }\href {https://doi.org/10.1103/PhysRevResearch.4.013113} {\bibfield  {journal} {\bibinfo  {journal} {Phys. Rev. Res.}\ }\textbf {\bibinfo {volume} {4}},\ \bibinfo {pages} {013113} (\bibinfo {year} {2022})}\BibitemShut {NoStop}%
\bibitem [{\citenamefont {Mishra}\ and\ \citenamefont {Bayat}(2021)}]{mishra2021driving}%
  \BibitemOpen
  \bibfield  {author} {\bibinfo {author} {\bibfnamefont {U.}~\bibnamefont {Mishra}}\ and\ \bibinfo {author} {\bibfnamefont {A.}~\bibnamefont {Bayat}},\ }\bibfield  {title} {\bibinfo {title} {Driving enhanced quantum sensing in partially accessible many-body systems},\ }\href {https://doi.org/10.1103/PhysRevLett.127.080504} {\bibfield  {journal} {\bibinfo  {journal} {Phys. Rev. Lett.}\ }\textbf {\bibinfo {volume} {127}},\ \bibinfo {pages} {080504} (\bibinfo {year} {2021})}\BibitemShut {NoStop}%
\bibitem [{\citenamefont {Mishra}\ and\ \citenamefont {Bayat}(2022)}]{mishra2022integrable}%
  \BibitemOpen
  \bibfield  {author} {\bibinfo {author} {\bibfnamefont {U.}~\bibnamefont {Mishra}}\ and\ \bibinfo {author} {\bibfnamefont {A.}~\bibnamefont {Bayat}},\ }\bibfield  {title} {\bibinfo {title} {Integrable quantum many-body sensors for {AC} field sensing},\ }\href {https://doi.org/10.1038/s41598-022-17381-y} {\bibfield  {journal} {\bibinfo  {journal} {Sci. Rep.}\ }\textbf {\bibinfo {volume} {12}},\ \bibinfo {pages} {14760} (\bibinfo {year} {2022})}\BibitemShut {NoStop}%
\bibitem [{\citenamefont {He}\ \emph {et~al.}(2023)\citenamefont {He}, \citenamefont {Yousefjani},\ and\ \citenamefont {Bayat}}]{he2023stark}%
  \BibitemOpen
  \bibfield  {author} {\bibinfo {author} {\bibfnamefont {X.}~\bibnamefont {He}}, \bibinfo {author} {\bibfnamefont {R.}~\bibnamefont {Yousefjani}},\ and\ \bibinfo {author} {\bibfnamefont {A.}~\bibnamefont {Bayat}},\ }\bibfield  {title} {\bibinfo {title} {{Stark} localization as a resource for weak-field sensing with super-{H}eisenberg precision},\ }\href {https://doi.org/10.1103/PhysRevLett.131.010801} {\bibfield  {journal} {\bibinfo  {journal} {Phys. Rev. Lett.}\ }\textbf {\bibinfo {volume} {131}},\ \bibinfo {pages} {010801} (\bibinfo {year} {2023})}\BibitemShut {NoStop}%
\bibitem [{\citenamefont {Yousefjani}\ \emph {et~al.}(2023{\natexlab{b}})\citenamefont {Yousefjani}, \citenamefont {He},\ and\ \citenamefont {Bayat}}]{yousefjani2023Long}%
  \BibitemOpen
  \bibfield  {author} {\bibinfo {author} {\bibfnamefont {R.}~\bibnamefont {Yousefjani}}, \bibinfo {author} {\bibfnamefont {X.}~\bibnamefont {He}},\ and\ \bibinfo {author} {\bibfnamefont {A.}~\bibnamefont {Bayat}},\ }\bibfield  {title} {\bibinfo {title} {Long-range interacting {Stark} many-body probes with super-{Heisenberg} precision},\ }\href {https://doi.org/10.1088/1674-1056/acf302} {\bibfield  {journal} {\bibinfo  {journal} {Chin. Phys. B}\ }\textbf {\bibinfo {volume} {32}},\ \bibinfo {pages} {100313} (\bibinfo {year} {2023}{\natexlab{b}})}\BibitemShut {NoStop}%
\bibitem [{\citenamefont {Yousefjani}\ \emph {et~al.}(2025{\natexlab{a}})\citenamefont {Yousefjani}, \citenamefont {He}, \citenamefont {Carollo},\ and\ \citenamefont {Bayat}}]{Yousefjani2025PRApplied}%
  \BibitemOpen
  \bibfield  {author} {\bibinfo {author} {\bibfnamefont {R.}~\bibnamefont {Yousefjani}}, \bibinfo {author} {\bibfnamefont {X.}~\bibnamefont {He}}, \bibinfo {author} {\bibfnamefont {A.}~\bibnamefont {Carollo}},\ and\ \bibinfo {author} {\bibfnamefont {A.}~\bibnamefont {Bayat}},\ }\bibfield  {title} {\bibinfo {title} {Nonlinearity-enhanced quantum sensing in stark probes},\ }\href {https://doi.org/10.1103/PhysRevApplied.23.014019} {\bibfield  {journal} {\bibinfo  {journal} {Phys. Rev. Applied}\ }\textbf {\bibinfo {volume} {23}},\ \bibinfo {pages} {014019} (\bibinfo {year} {2025}{\natexlab{a}})}\BibitemShut {NoStop}%
\bibitem [{\citenamefont {Yousefjani}\ and\ \citenamefont {Al-Kuwari}(2026)}]{yousefjani2026exponentially}%
  \BibitemOpen
  \bibfield  {author} {\bibinfo {author} {\bibfnamefont {R.}~\bibnamefont {Yousefjani}}\ and\ \bibinfo {author} {\bibfnamefont {S.}~\bibnamefont {Al-Kuwari}},\ }\href {https://arxiv.org/abs/2604.17262} {\bibinfo {title} {Exponentially-enhanced weak-field sensing with quantum stark localization}} (\bibinfo {year} {2026}),\ \Eprint {https://arxiv.org/abs/2604.17262} {arXiv:2604.17262 [quant-ph]} \BibitemShut {NoStop}%
\bibitem [{\citenamefont {Gyhm}\ \emph {et~al.}(2025)\citenamefont {Gyhm}, \citenamefont {Kwon},\ and\ \citenamefont {Hwang}}]{gyhm2025fundamental}%
  \BibitemOpen
  \bibfield  {author} {\bibinfo {author} {\bibfnamefont {J.-Y.}\ \bibnamefont {Gyhm}}, \bibinfo {author} {\bibfnamefont {H.}~\bibnamefont {Kwon}},\ and\ \bibinfo {author} {\bibfnamefont {M.-J.}\ \bibnamefont {Hwang}},\ }\href {https://arxiv.org/abs/2506.19003} {\bibinfo {title} {Fundamental scaling limit in critical quantum metrology}} (\bibinfo {year} {2025}),\ \Eprint {https://arxiv.org/abs/2506.19003} {arXiv:2506.19003 [quant-ph]} \BibitemShut {NoStop}%
\bibitem [{\citenamefont {Sarkar}\ \emph {et~al.}(2025)\citenamefont {Sarkar}, \citenamefont {Bayat}, \citenamefont {Bose},\ and\ \citenamefont {Ghosh}}]{sarkar2025exponentially}%
  \BibitemOpen
  \bibfield  {author} {\bibinfo {author} {\bibfnamefont {S.}~\bibnamefont {Sarkar}}, \bibinfo {author} {\bibfnamefont {A.}~\bibnamefont {Bayat}}, \bibinfo {author} {\bibfnamefont {S.}~\bibnamefont {Bose}},\ and\ \bibinfo {author} {\bibfnamefont {R.}~\bibnamefont {Ghosh}},\ }\bibfield  {title} {\bibinfo {title} {Exponentially-enhanced quantum sensing with many-body phase transitions},\ }\href {https://doi.org/10.1038/s41467-025-60291-6} {\bibfield  {journal} {\bibinfo  {journal} {Nat. Commun.}\ }\textbf {\bibinfo {volume} {16}},\ \bibinfo {pages} {5159} (\bibinfo {year} {2025})}\BibitemShut {NoStop}%
\bibitem [{\citenamefont {Liu}\ \emph {et~al.}(2021)\citenamefont {Liu}, \citenamefont {Chen}, \citenamefont {Jiang}, \citenamefont {Yang}, \citenamefont {Wu}, \citenamefont {Huan}, \citenamefont {Fei}, \citenamefont {Gong}, \citenamefont {Su}, \citenamefont {Lan}, \citenamefont {Lu}, \citenamefont {Yuan},\ and\ \citenamefont {Wang}}]{Liu2021ExperimentalCritical}%
  \BibitemOpen
  \bibfield  {author} {\bibinfo {author} {\bibfnamefont {R.}~\bibnamefont {Liu}}, \bibinfo {author} {\bibfnamefont {Y.}~\bibnamefont {Chen}}, \bibinfo {author} {\bibfnamefont {M.}~\bibnamefont {Jiang}}, \bibinfo {author} {\bibfnamefont {X.}~\bibnamefont {Yang}}, \bibinfo {author} {\bibfnamefont {Z.}~\bibnamefont {Wu}}, \bibinfo {author} {\bibfnamefont {Y.}~\bibnamefont {Huan}}, \bibinfo {author} {\bibfnamefont {J.}~\bibnamefont {Fei}}, \bibinfo {author} {\bibfnamefont {X.}~\bibnamefont {Gong}}, \bibinfo {author} {\bibfnamefont {Z.}~\bibnamefont {Su}}, \bibinfo {author} {\bibfnamefont {T.}~\bibnamefont {Lan}}, \bibinfo {author} {\bibfnamefont {J.}~\bibnamefont {Lu}}, \bibinfo {author} {\bibfnamefont {H.}~\bibnamefont {Yuan}},\ and\ \bibinfo {author} {\bibfnamefont {X.}~\bibnamefont {Wang}},\ }\bibfield  {title} {\bibinfo {title} {Experimental critical quantum metrology with the heisenberg scaling},\ }\href {https://doi.org/10.1038/s41534-021-00507-x} {\bibfield  {journal} {\bibinfo  {journal} {npj Quantum
  Inf.}\ }\textbf {\bibinfo {volume} {7}},\ \bibinfo {pages} {170} (\bibinfo {year} {2021})}\BibitemShut {NoStop}%
\bibitem [{\citenamefont {Ding}\ \emph {et~al.}(2022)\citenamefont {Ding}, \citenamefont {Liu}, \citenamefont {Shi}, \citenamefont {Guo}, \citenamefont {M{\o}lmer},\ and\ \citenamefont {Adams}}]{Ding2022EnhancedMetrology}%
  \BibitemOpen
  \bibfield  {author} {\bibinfo {author} {\bibfnamefont {D.-S.}\ \bibnamefont {Ding}}, \bibinfo {author} {\bibfnamefont {Z.-K.}\ \bibnamefont {Liu}}, \bibinfo {author} {\bibfnamefont {B.-S.}\ \bibnamefont {Shi}}, \bibinfo {author} {\bibfnamefont {G.-C.}\ \bibnamefont {Guo}}, \bibinfo {author} {\bibfnamefont {K.}~\bibnamefont {M{\o}lmer}},\ and\ \bibinfo {author} {\bibfnamefont {C.~S.}\ \bibnamefont {Adams}},\ }\bibfield  {title} {\bibinfo {title} {Enhanced metrology at the critical point of a many-body rydberg atomic system},\ }\href {https://doi.org/10.1038/s41567-022-01777-8} {\bibfield  {journal} {\bibinfo  {journal} {Nat. Phys.}\ }\textbf {\bibinfo {volume} {18}},\ \bibinfo {pages} {1447} (\bibinfo {year} {2022})}\BibitemShut {NoStop}%
\bibitem [{\citenamefont {Yu}\ \emph {et~al.}(2025)\citenamefont {Yu}, \citenamefont {Liu}, \citenamefont {Xue}, \citenamefont {Yang}, \citenamefont {Wang}, \citenamefont {Zhang}, \citenamefont {Cui}, \citenamefont {Yang}, \citenamefont {Li}, \citenamefont {Han},\ and\ \citenamefont {Yu}}]{yu2025experimental}%
  \BibitemOpen
  \bibfield  {author} {\bibinfo {author} {\bibfnamefont {Y.}~\bibnamefont {Yu}}, \bibinfo {author} {\bibfnamefont {R.}~\bibnamefont {Liu}}, \bibinfo {author} {\bibfnamefont {G.}~\bibnamefont {Xue}}, \bibinfo {author} {\bibfnamefont {C.}~\bibnamefont {Yang}}, \bibinfo {author} {\bibfnamefont {C.}~\bibnamefont {Wang}}, \bibinfo {author} {\bibfnamefont {J.}~\bibnamefont {Zhang}}, \bibinfo {author} {\bibfnamefont {J.}~\bibnamefont {Cui}}, \bibinfo {author} {\bibfnamefont {X.}~\bibnamefont {Yang}}, \bibinfo {author} {\bibfnamefont {J.}~\bibnamefont {Li}}, \bibinfo {author} {\bibfnamefont {J.}~\bibnamefont {Han}},\ and\ \bibinfo {author} {\bibfnamefont {H.}~\bibnamefont {Yu}},\ }\href {https://arxiv.org/abs/2501.04955} {\bibinfo {title} {Experimental realization of criticality-enhanced global quantum sensing via non-equilibrium dynamics}} (\bibinfo {year} {2025}),\ \Eprint {https://arxiv.org/abs/2501.04955} {arXiv:2501.04955 [quant-ph]} \BibitemShut {NoStop}%
\bibitem [{\citenamefont {Li}\ \emph {et~al.}(2025)\citenamefont {Li}, \citenamefont {Yang}, \citenamefont {Shi}, \citenamefont {Wang}, \citenamefont {Wang}, \citenamefont {Li}, \citenamefont {Zhang}, \citenamefont {Zhao}, \citenamefont {Xu}, \citenamefont {Deng}, \citenamefont {Yu}, \citenamefont {Ma}, \citenamefont {Li}, \citenamefont {Zhang}, \citenamefont {Fang}, \citenamefont {Song}, \citenamefont {Liu}, \citenamefont {Zhou}, \citenamefont {Liu}, \citenamefont {Chen}, \citenamefont {Liang}, \citenamefont {Song}, \citenamefont {Xiang}, \citenamefont {Xu}, \citenamefont {Huang}, \citenamefont {Bayat},\ and\ \citenamefont {Fan}}]{li2025non}%
  \BibitemOpen
  \bibfield  {author} {\bibinfo {author} {\bibfnamefont {H.}~\bibnamefont {Li}}, \bibinfo {author} {\bibfnamefont {Y.}~\bibnamefont {Yang}}, \bibinfo {author} {\bibfnamefont {Y.-H.}\ \bibnamefont {Shi}}, \bibinfo {author} {\bibfnamefont {Z.-A.}\ \bibnamefont {Wang}}, \bibinfo {author} {\bibfnamefont {Z.}~\bibnamefont {Wang}}, \bibinfo {author} {\bibfnamefont {J.}~\bibnamefont {Li}}, \bibinfo {author} {\bibfnamefont {Y.}~\bibnamefont {Zhang}}, \bibinfo {author} {\bibfnamefont {K.}~\bibnamefont {Zhao}}, \bibinfo {author} {\bibfnamefont {Y.-S.}\ \bibnamefont {Xu}}, \bibinfo {author} {\bibfnamefont {C.-L.}\ \bibnamefont {Deng}}, \bibinfo {author} {\bibfnamefont {L.}~\bibnamefont {Yu}}, \bibinfo {author} {\bibfnamefont {W.-G.}\ \bibnamefont {Ma}}, \bibinfo {author} {\bibfnamefont {T.-M.}\ \bibnamefont {Li}}, \bibinfo {author} {\bibfnamefont {J.-C.}\ \bibnamefont {Zhang}}, \bibinfo {author} {\bibfnamefont {C.-P.}\ \bibnamefont {Fang}}, \bibinfo {author} {\bibfnamefont {J.-C.}\ \bibnamefont {Song}}, \bibinfo
  {author} {\bibfnamefont {H.-T.}\ \bibnamefont {Liu}}, \bibinfo {author} {\bibfnamefont {S.-Y.}\ \bibnamefont {Zhou}}, \bibinfo {author} {\bibfnamefont {Z.-H.}\ \bibnamefont {Liu}}, \bibinfo {author} {\bibfnamefont {B.-J.}\ \bibnamefont {Chen}}, \bibinfo {author} {\bibfnamefont {G.-H.}\ \bibnamefont {Liang}}, \bibinfo {author} {\bibfnamefont {X.}~\bibnamefont {Song}}, \bibinfo {author} {\bibfnamefont {Z.}~\bibnamefont {Xiang}}, \bibinfo {author} {\bibfnamefont {K.}~\bibnamefont {Xu}}, \bibinfo {author} {\bibfnamefont {K.}~\bibnamefont {Huang}}, \bibinfo {author} {\bibfnamefont {A.}~\bibnamefont {Bayat}},\ and\ \bibinfo {author} {\bibfnamefont {H.}~\bibnamefont {Fan}},\ }\href {https://arxiv.org/abs/2508.14409} {\bibinfo {title} {Non-equilibrium criticality-enhanced quantum sensing with superconducting qubits}} (\bibinfo {year} {2025}),\ \Eprint {https://arxiv.org/abs/2508.14409} {arXiv:2508.14409 [quant-ph]} \BibitemShut {NoStop}%
\bibitem [{\citenamefont {Xiao}\ \emph {et~al.}(2026)\citenamefont {Xiao}, \citenamefont {Sarkar}, \citenamefont {Wang}, \citenamefont {Bayat},\ and\ \citenamefont {Xue}}]{Xiao2026CriticalityEnhancedQuantumSensingWalks}%
  \BibitemOpen
  \bibfield  {author} {\bibinfo {author} {\bibfnamefont {L.}~\bibnamefont {Xiao}}, \bibinfo {author} {\bibfnamefont {S.}~\bibnamefont {Sarkar}}, \bibinfo {author} {\bibfnamefont {K.}~\bibnamefont {Wang}}, \bibinfo {author} {\bibfnamefont {A.}~\bibnamefont {Bayat}},\ and\ \bibinfo {author} {\bibfnamefont {P.}~\bibnamefont {Xue}},\ }\bibfield  {title} {\bibinfo {title} {Observation of criticality-enhanced quantum sensing in nonunitary quantum walks},\ }\href {https://doi.org/10.1103/6gql-zgkb} {\bibfield  {journal} {\bibinfo  {journal} {Phys. Rev. Lett.}\ }\textbf {\bibinfo {volume} {136}},\ \bibinfo {pages} {060802} (\bibinfo {year} {2026})}\BibitemShut {NoStop}%
\bibitem [{\citenamefont {Xiao}\ \emph {et~al.}(2024)\citenamefont {Xiao}, \citenamefont {Chu}, \citenamefont {Lin}, \citenamefont {Lin}, \citenamefont {Yi}, \citenamefont {Cai},\ and\ \citenamefont {Xue}}]{Xiao2024NonHermitianSensingNoEP}%
  \BibitemOpen
  \bibfield  {author} {\bibinfo {author} {\bibfnamefont {L.}~\bibnamefont {Xiao}}, \bibinfo {author} {\bibfnamefont {Y.}~\bibnamefont {Chu}}, \bibinfo {author} {\bibfnamefont {Q.}~\bibnamefont {Lin}}, \bibinfo {author} {\bibfnamefont {H.}~\bibnamefont {Lin}}, \bibinfo {author} {\bibfnamefont {W.}~\bibnamefont {Yi}}, \bibinfo {author} {\bibfnamefont {J.}~\bibnamefont {Cai}},\ and\ \bibinfo {author} {\bibfnamefont {P.}~\bibnamefont {Xue}},\ }\bibfield  {title} {\bibinfo {title} {Non-hermitian sensing in the absence of exceptional points},\ }\href {https://doi.org/10.1103/PhysRevLett.133.180801} {\bibfield  {journal} {\bibinfo  {journal} {Phys. Rev. Lett.}\ }\textbf {\bibinfo {volume} {133}},\ \bibinfo {pages} {180801} (\bibinfo {year} {2024})}\BibitemShut {NoStop}%
\bibitem [{\citenamefont {Tong}\ \emph {et~al.}(2025)\citenamefont {Tong}, \citenamefont {Qiu}, \citenamefont {Zhan}, \citenamefont {Lin}, \citenamefont {Wang}, \citenamefont {Nori},\ and\ \citenamefont {Xue}}]{Tong2025TopologicalSensingQuantumWalkDefects}%
  \BibitemOpen
  \bibfield  {author} {\bibinfo {author} {\bibfnamefont {X.}~\bibnamefont {Tong}}, \bibinfo {author} {\bibfnamefont {X.}~\bibnamefont {Qiu}}, \bibinfo {author} {\bibfnamefont {X.}~\bibnamefont {Zhan}}, \bibinfo {author} {\bibfnamefont {Q.}~\bibnamefont {Lin}}, \bibinfo {author} {\bibfnamefont {K.}~\bibnamefont {Wang}}, \bibinfo {author} {\bibfnamefont {F.}~\bibnamefont {Nori}},\ and\ \bibinfo {author} {\bibfnamefont {P.}~\bibnamefont {Xue}},\ }\bibfield  {title} {\bibinfo {title} {Topological sensing in the dynamics of quantum walks with defects},\ }\href {https://doi.org/10.1103/3jtm-nxt4} {\bibfield  {journal} {\bibinfo  {journal} {Phys. Rev. B}\ }\textbf {\bibinfo {volume} {112}},\ \bibinfo {pages} {L241116} (\bibinfo {year} {2025})}\BibitemShut {NoStop}%
\bibitem [{\citenamefont {Beaulieu}\ \emph {et~al.}(2025)\citenamefont {Beaulieu}, \citenamefont {Minganti}, \citenamefont {Frasca}, \citenamefont {Scigliuzzo}, \citenamefont {Felicetti}, \citenamefont {Di~Candia},\ and\ \citenamefont {Scarlino}}]{Beaulieu2025CriticalityEnhancedParametricResonator}%
  \BibitemOpen
  \bibfield  {author} {\bibinfo {author} {\bibfnamefont {G.}~\bibnamefont {Beaulieu}}, \bibinfo {author} {\bibfnamefont {F.}~\bibnamefont {Minganti}}, \bibinfo {author} {\bibfnamefont {S.}~\bibnamefont {Frasca}}, \bibinfo {author} {\bibfnamefont {M.}~\bibnamefont {Scigliuzzo}}, \bibinfo {author} {\bibfnamefont {S.}~\bibnamefont {Felicetti}}, \bibinfo {author} {\bibfnamefont {R.}~\bibnamefont {Di~Candia}},\ and\ \bibinfo {author} {\bibfnamefont {P.}~\bibnamefont {Scarlino}},\ }\bibfield  {title} {\bibinfo {title} {Criticality-enhanced quantum sensing with a parametric superconducting resonator},\ }\href {https://doi.org/10.1103/PRXQuantum.6.020301} {\bibfield  {journal} {\bibinfo  {journal} {PRX Quantum}\ }\textbf {\bibinfo {volume} {6}},\ \bibinfo {pages} {020301} (\bibinfo {year} {2025})}\BibitemShut {NoStop}%
\bibitem [{\citenamefont {Montenegro}\ \emph {et~al.}(2025)\citenamefont {Montenegro}, \citenamefont {Mukhopadhyay}, \citenamefont {Yousefjani}, \citenamefont {Sarkar}, \citenamefont {Mishra}, \citenamefont {Paris},\ and\ \citenamefont {Bayat}}]{Montenegro2025Review}%
  \BibitemOpen
  \bibfield  {author} {\bibinfo {author} {\bibfnamefont {V.}~\bibnamefont {Montenegro}}, \bibinfo {author} {\bibfnamefont {C.}~\bibnamefont {Mukhopadhyay}}, \bibinfo {author} {\bibfnamefont {R.}~\bibnamefont {Yousefjani}}, \bibinfo {author} {\bibfnamefont {S.}~\bibnamefont {Sarkar}}, \bibinfo {author} {\bibfnamefont {U.}~\bibnamefont {Mishra}}, \bibinfo {author} {\bibfnamefont {M.~G.~A.}\ \bibnamefont {Paris}},\ and\ \bibinfo {author} {\bibfnamefont {A.}~\bibnamefont {Bayat}},\ }\bibfield  {title} {\bibinfo {title} {Review: Quantum metrology and sensing with many-body systems},\ }\href {https://doi.org/10.1016/j.physrep.2025.05.005} {\bibfield  {journal} {\bibinfo  {journal} {Physics Reports}\ }\textbf {\bibinfo {volume} {1134}},\ \bibinfo {pages} {1} (\bibinfo {year} {2025})}\BibitemShut {NoStop}%
\bibitem [{\citenamefont {Lyu}\ \emph {et~al.}(2020)\citenamefont {Lyu}, \citenamefont {Choudhury}, \citenamefont {Lv}, \citenamefont {Yan},\ and\ \citenamefont {Zhou}}]{lyu2020eternal}%
  \BibitemOpen
  \bibfield  {author} {\bibinfo {author} {\bibfnamefont {C.}~\bibnamefont {Lyu}}, \bibinfo {author} {\bibfnamefont {S.}~\bibnamefont {Choudhury}}, \bibinfo {author} {\bibfnamefont {C.}~\bibnamefont {Lv}}, \bibinfo {author} {\bibfnamefont {Y.}~\bibnamefont {Yan}},\ and\ \bibinfo {author} {\bibfnamefont {Q.}~\bibnamefont {Zhou}},\ }\bibfield  {title} {\bibinfo {title} {Eternal discrete time crystal beating the heisenberg limit},\ }\href {https://doi.org/10.1103/PhysRevResearch.2.033070} {\bibfield  {journal} {\bibinfo  {journal} {Phys. Rev. Res.}\ }\textbf {\bibinfo {volume} {2}},\ \bibinfo {pages} {033070} (\bibinfo {year} {2020})}\BibitemShut {NoStop}%
\bibitem [{\citenamefont {Iemini}\ \emph {et~al.}(2024)\citenamefont {Iemini}, \citenamefont {Fazio},\ and\ \citenamefont {Sanpera}}]{iemini2023floquet}%
  \BibitemOpen
  \bibfield  {author} {\bibinfo {author} {\bibfnamefont {F.}~\bibnamefont {Iemini}}, \bibinfo {author} {\bibfnamefont {R.}~\bibnamefont {Fazio}},\ and\ \bibinfo {author} {\bibfnamefont {A.}~\bibnamefont {Sanpera}},\ }\bibfield  {title} {\bibinfo {title} {Floquet time crystals as quantum sensors of {AC} fields},\ }\href {https://doi.org/10.1103/PhysRevA.109.L050203} {\bibfield  {journal} {\bibinfo  {journal} {Phys. Rev. A}\ }\textbf {\bibinfo {volume} {109}},\ \bibinfo {pages} {L050203} (\bibinfo {year} {2024})}\BibitemShut {NoStop}%
\bibitem [{\citenamefont {Yousefjani}\ \emph {et~al.}(2025{\natexlab{b}})\citenamefont {Yousefjani}, \citenamefont {Sacha},\ and\ \citenamefont {Bayat}}]{yousefjani2025discrete}%
  \BibitemOpen
  \bibfield  {author} {\bibinfo {author} {\bibfnamefont {R.}~\bibnamefont {Yousefjani}}, \bibinfo {author} {\bibfnamefont {K.}~\bibnamefont {Sacha}},\ and\ \bibinfo {author} {\bibfnamefont {A.}~\bibnamefont {Bayat}},\ }\bibfield  {title} {\bibinfo {title} {Discrete time crystal phase as a resource for quantum-enhanced sensing},\ }\href {https://doi.org/10.1103/PhysRevB.111.125159} {\bibfield  {journal} {\bibinfo  {journal} {Phys. Rev. B}\ }\textbf {\bibinfo {volume} {111}},\ \bibinfo {pages} {125159} (\bibinfo {year} {2025}{\natexlab{b}})}\BibitemShut {NoStop}%
\bibitem [{\citenamefont {Yousefjani}\ \emph {et~al.}(2025{\natexlab{c}})\citenamefont {Yousefjani}, \citenamefont {Al-Kuwari},\ and\ \citenamefont {Bayat}}]{Yousefjani2025PeriodicFieldDTC}%
  \BibitemOpen
  \bibfield  {author} {\bibinfo {author} {\bibfnamefont {R.}~\bibnamefont {Yousefjani}}, \bibinfo {author} {\bibfnamefont {S.}~\bibnamefont {Al-Kuwari}},\ and\ \bibinfo {author} {\bibfnamefont {A.}~\bibnamefont {Bayat}},\ }\bibfield  {title} {\bibinfo {title} {Discrete time crystal for periodic-field sensing with quantum-enhanced precision},\ }\href {https://doi.org/10.1103/7m63-lnb8} {\bibfield  {journal} {\bibinfo  {journal} {Phys. Rev. Applied}\ }\textbf {\bibinfo {volume} {24}},\ \bibinfo {pages} {054047} (\bibinfo {year} {2025}{\natexlab{c}})}\BibitemShut {NoStop}%
\bibitem [{\citenamefont {Tsypilnikov}\ \emph {et~al.}(2026)\citenamefont {Tsypilnikov}, \citenamefont {Fibger},\ and\ \citenamefont {Iemini}}]{tsypilnikov2026exact}%
  \BibitemOpen
  \bibfield  {author} {\bibinfo {author} {\bibfnamefont {A.}~\bibnamefont {Tsypilnikov}}, \bibinfo {author} {\bibfnamefont {M.}~\bibnamefont {Fibger}},\ and\ \bibinfo {author} {\bibfnamefont {F.}~\bibnamefont {Iemini}},\ }\bibfield  {title} {\bibinfo {title} {Exact analysis of {AC} sensors based on floquet time crystals},\ }\href {https://doi.org/10.1103/9h67-kqyz} {\bibfield  {journal} {\bibinfo  {journal} {Phys. Rev. A}\ }\textbf {\bibinfo {volume} {113}},\ \bibinfo {pages} {022620} (\bibinfo {year} {2026})}\BibitemShut {NoStop}%
\bibitem [{\citenamefont {Shukla}\ \emph {et~al.}(2025)\citenamefont {Shukla}, \citenamefont {Chotorlishvili}, \citenamefont {Mishra},\ and\ \citenamefont {Iemini}}]{shukla2025prethermal}%
  \BibitemOpen
  \bibfield  {author} {\bibinfo {author} {\bibfnamefont {R.~K.}\ \bibnamefont {Shukla}}, \bibinfo {author} {\bibfnamefont {L.}~\bibnamefont {Chotorlishvili}}, \bibinfo {author} {\bibfnamefont {S.~K.}\ \bibnamefont {Mishra}},\ and\ \bibinfo {author} {\bibfnamefont {F.}~\bibnamefont {Iemini}},\ }\bibfield  {title} {\bibinfo {title} {Prethermal floquet time crystals in chiral multiferroic chains and applications as quantum sensors of {AC} fields},\ }\href {https://doi.org/10.1103/PhysRevB.111.024315} {\bibfield  {journal} {\bibinfo  {journal} {Phys. Rev. B}\ }\textbf {\bibinfo {volume} {111}},\ \bibinfo {pages} {024315} (\bibinfo {year} {2025})}\BibitemShut {NoStop}%
\bibitem [{\citenamefont {Montenegro}\ \emph {et~al.}(2023)\citenamefont {Montenegro}, \citenamefont {Cejnar}, \citenamefont {Gietka},\ and\ \citenamefont {Carollo}}]{montenegro2023quantum}%
  \BibitemOpen
  \bibfield  {author} {\bibinfo {author} {\bibfnamefont {V.}~\bibnamefont {Montenegro}}, \bibinfo {author} {\bibfnamefont {P.}~\bibnamefont {Cejnar}}, \bibinfo {author} {\bibfnamefont {K.}~\bibnamefont {Gietka}},\ and\ \bibinfo {author} {\bibfnamefont {A.}~\bibnamefont {Carollo}},\ }\bibfield  {title} {\bibinfo {title} {Quantum metrology with boundary time crystals},\ }\href {https://doi.org/10.1038/s42005-023-01423-6} {\bibfield  {journal} {\bibinfo  {journal} {Commun. Phys.}\ }\textbf {\bibinfo {volume} {6}},\ \bibinfo {pages} {304} (\bibinfo {year} {2023})}\BibitemShut {NoStop}%
\bibitem [{\citenamefont {Cabot}\ \emph {et~al.}(2024)\citenamefont {Cabot}, \citenamefont {Carollo},\ and\ \citenamefont {Lesanovsky}}]{Cabot2024Continuous}%
  \BibitemOpen
  \bibfield  {author} {\bibinfo {author} {\bibfnamefont {A.}~\bibnamefont {Cabot}}, \bibinfo {author} {\bibfnamefont {A.}~\bibnamefont {Carollo}},\ and\ \bibinfo {author} {\bibfnamefont {I.}~\bibnamefont {Lesanovsky}},\ }\bibfield  {title} {\bibinfo {title} {Continuous sensing and parameter estimation with the boundary time crystal},\ }\href {https://doi.org/10.1103/PhysRevLett.132.050801} {\bibfield  {journal} {\bibinfo  {journal} {Phys. Rev. Lett.}\ }\textbf {\bibinfo {volume} {132}},\ \bibinfo {pages} {050801} (\bibinfo {year} {2024})}\BibitemShut {NoStop}%
\bibitem [{\citenamefont {Viotti}\ \emph {et~al.}(2026)\citenamefont {Viotti}, \citenamefont {Huber}, \citenamefont {Fazio},\ and\ \citenamefont {Manzano}}]{Viotti2026QuantumTimeCrystalClock}%
  \BibitemOpen
  \bibfield  {author} {\bibinfo {author} {\bibfnamefont {L.}~\bibnamefont {Viotti}}, \bibinfo {author} {\bibfnamefont {M.}~\bibnamefont {Huber}}, \bibinfo {author} {\bibfnamefont {R.}~\bibnamefont {Fazio}},\ and\ \bibinfo {author} {\bibfnamefont {G.}~\bibnamefont {Manzano}},\ }\bibfield  {title} {\bibinfo {title} {Quantum time crystal clock and its performance},\ }\href {https://doi.org/10.1103/dj21-gmdj} {\bibfield  {journal} {\bibinfo  {journal} {Phys. Rev. Lett.}\ }\textbf {\bibinfo {volume} {136}},\ \bibinfo {pages} {110401} (\bibinfo {year} {2026})}\BibitemShut {NoStop}%
\bibitem [{\citenamefont {Gribben}\ \emph {et~al.}(2025)\citenamefont {Gribben}, \citenamefont {Sanpera}, \citenamefont {Fazio}, \citenamefont {Marino},\ and\ \citenamefont {Iemini}}]{Gribben2025BoundaryTimeCrystalsACSensors}%
  \BibitemOpen
  \bibfield  {author} {\bibinfo {author} {\bibfnamefont {D.}~\bibnamefont {Gribben}}, \bibinfo {author} {\bibfnamefont {A.}~\bibnamefont {Sanpera}}, \bibinfo {author} {\bibfnamefont {R.}~\bibnamefont {Fazio}}, \bibinfo {author} {\bibfnamefont {J.}~\bibnamefont {Marino}},\ and\ \bibinfo {author} {\bibfnamefont {F.}~\bibnamefont {Iemini}},\ }\bibfield  {title} {\bibinfo {title} {Boundary time crystals as ac sensors: Enhancements and constraints},\ }\href {https://doi.org/10.21468/SciPostPhys.18.3.100} {\bibfield  {journal} {\bibinfo  {journal} {SciPost Physics}\ }\textbf {\bibinfo {volume} {18}},\ \bibinfo {pages} {100} (\bibinfo {year} {2025})}\BibitemShut {NoStop}%
\bibitem [{\citenamefont {Sierant}\ \emph {et~al.}(2022)\citenamefont {Sierant}, \citenamefont {Chiriac{\`o}}, \citenamefont {Surace}, \citenamefont {Sharma}, \citenamefont {Turkeshi}, \citenamefont {Dalmonte}, \citenamefont {Fazio},\ and\ \citenamefont {Pagano}}]{Sierant2022DissipativeFloquetDynamics}%
  \BibitemOpen
  \bibfield  {author} {\bibinfo {author} {\bibfnamefont {P.}~\bibnamefont {Sierant}}, \bibinfo {author} {\bibfnamefont {G.}~\bibnamefont {Chiriac{\`o}}}, \bibinfo {author} {\bibfnamefont {F.~M.}\ \bibnamefont {Surace}}, \bibinfo {author} {\bibfnamefont {S.}~\bibnamefont {Sharma}}, \bibinfo {author} {\bibfnamefont {X.}~\bibnamefont {Turkeshi}}, \bibinfo {author} {\bibfnamefont {M.}~\bibnamefont {Dalmonte}}, \bibinfo {author} {\bibfnamefont {R.}~\bibnamefont {Fazio}},\ and\ \bibinfo {author} {\bibfnamefont {G.}~\bibnamefont {Pagano}},\ }\bibfield  {title} {\bibinfo {title} {Dissipative floquet dynamics: From steady state to measurement induced criticality in trapped-ion chains},\ }\href {https://doi.org/10.22331/q-2022-02-02-638} {\bibfield  {journal} {\bibinfo  {journal} {Quantum}\ }\textbf {\bibinfo {volume} {6}},\ \bibinfo {pages} {638} (\bibinfo {year} {2022})}\BibitemShut {NoStop}%
\bibitem [{\citenamefont {Moon}\ \emph {et~al.}(2026)\citenamefont {Moon}, \citenamefont {Schindler}, \citenamefont {Smith}, \citenamefont {Druga}, \citenamefont {Zhang}, \citenamefont {Bukov},\ and\ \citenamefont {Ajoy}}]{Moon2026DTCsensing}%
  \BibitemOpen
  \bibfield  {author} {\bibinfo {author} {\bibfnamefont {L.~J.~I.}\ \bibnamefont {Moon}}, \bibinfo {author} {\bibfnamefont {P.~M.}\ \bibnamefont {Schindler}}, \bibinfo {author} {\bibfnamefont {R.~J.}\ \bibnamefont {Smith}}, \bibinfo {author} {\bibfnamefont {E.}~\bibnamefont {Druga}}, \bibinfo {author} {\bibfnamefont {Z.-R.}\ \bibnamefont {Zhang}}, \bibinfo {author} {\bibfnamefont {M.}~\bibnamefont {Bukov}},\ and\ \bibinfo {author} {\bibfnamefont {A.}~\bibnamefont {Ajoy}},\ }\bibfield  {title} {\bibinfo {title} {Sensing with discrete time crystals},\ }\href {https://doi.org/10.1038/s41567-025-03163-6} {\bibfield  {journal} {\bibinfo  {journal} {Nat. Phys.}\ }\textbf {\bibinfo {volume} {22}},\ \bibinfo {pages} {367} (\bibinfo {year} {2026})}\BibitemShut {NoStop}%
\bibitem [{\citenamefont {Fisher}(1922)}]{fisher1922mathematical}%
  \BibitemOpen
  \bibfield  {author} {\bibinfo {author} {\bibfnamefont {R.~A.}\ \bibnamefont {Fisher}},\ }\bibfield  {title} {\bibinfo {title} {On the mathematical foundations of theoretical statistics},\ }\href {https://doi.org/10.1098/rsta.1922.0009} {\bibfield  {journal} {\bibinfo  {journal} {Philos. Trans. R. Soc. Lond. A}\ }\textbf {\bibinfo {volume} {222}},\ \bibinfo {pages} {309} (\bibinfo {year} {1922})}\BibitemShut {NoStop}%
\bibitem [{\citenamefont {Cram{\'e}r}(1999)}]{cramer1999mathematical}%
  \BibitemOpen
  \bibfield  {author} {\bibinfo {author} {\bibfnamefont {H.}~\bibnamefont {Cram{\'e}r}},\ }\href {https://press.princeton.edu/books/paperback/9780691005478/mathematical-methods-of-statistics} {\emph {\bibinfo {title} {Mathematical Methods of Statistics}}}\ (\bibinfo  {publisher} {Princeton University Press},\ \bibinfo {year} {1999})\BibitemShut {NoStop}%
\bibitem [{\citenamefont {Braunstein}\ and\ \citenamefont {Caves}(1994)}]{braunstein1994statistical}%
  \BibitemOpen
  \bibfield  {author} {\bibinfo {author} {\bibfnamefont {S.~L.}\ \bibnamefont {Braunstein}}\ and\ \bibinfo {author} {\bibfnamefont {C.~M.}\ \bibnamefont {Caves}},\ }\bibfield  {title} {\bibinfo {title} {Statistical distance and the geometry of quantum states},\ }\href {https://doi.org/10.1103/PhysRevLett.72.3439} {\bibfield  {journal} {\bibinfo  {journal} {Phys. Rev. Lett.}\ }\textbf {\bibinfo {volume} {72}},\ \bibinfo {pages} {3439} (\bibinfo {year} {1994})}\BibitemShut {NoStop}%
\bibitem [{\citenamefont {Boixo}\ \emph {et~al.}(2007)\citenamefont {Boixo}, \citenamefont {Flammia}, \citenamefont {Caves},\ and\ \citenamefont {Geremia}}]{Boixo2007GeneralizedLimits}%
  \BibitemOpen
  \bibfield  {author} {\bibinfo {author} {\bibfnamefont {S.}~\bibnamefont {Boixo}}, \bibinfo {author} {\bibfnamefont {S.~T.}\ \bibnamefont {Flammia}}, \bibinfo {author} {\bibfnamefont {C.~M.}\ \bibnamefont {Caves}},\ and\ \bibinfo {author} {\bibfnamefont {J.~M.}\ \bibnamefont {Geremia}},\ }\bibfield  {title} {\bibinfo {title} {Generalized limits for single-parameter quantum estimation},\ }\href {https://doi.org/10.1103/PhysRevLett.98.090401} {\bibfield  {journal} {\bibinfo  {journal} {Phys. Rev. Lett.}\ }\textbf {\bibinfo {volume} {98}},\ \bibinfo {pages} {090401} (\bibinfo {year} {2007})}\BibitemShut {NoStop}%
\bibitem [{\citenamefont {Abiuso}\ \emph {et~al.}(2025)\citenamefont {Abiuso}, \citenamefont {Sekatski}, \citenamefont {Calsamiglia},\ and\ \citenamefont {Perarnau-Llobet}}]{Abiuso2025FundamentalLimitsThermalEquilibrium}%
  \BibitemOpen
  \bibfield  {author} {\bibinfo {author} {\bibfnamefont {P.}~\bibnamefont {Abiuso}}, \bibinfo {author} {\bibfnamefont {P.}~\bibnamefont {Sekatski}}, \bibinfo {author} {\bibfnamefont {J.}~\bibnamefont {Calsamiglia}},\ and\ \bibinfo {author} {\bibfnamefont {M.}~\bibnamefont {Perarnau-Llobet}},\ }\bibfield  {title} {\bibinfo {title} {Fundamental limits of metrology at thermal equilibrium},\ }\href {https://doi.org/10.1103/PhysRevLett.134.010801} {\bibfield  {journal} {\bibinfo  {journal} {Phys. Rev. Lett.}\ }\textbf {\bibinfo {volume} {134}},\ \bibinfo {pages} {010801} (\bibinfo {year} {2025})}\BibitemShut {NoStop}%
\bibitem [{\citenamefont {Puig}\ \emph {et~al.}(2025)\citenamefont {Puig}, \citenamefont {Sekatski}, \citenamefont {Erdman}, \citenamefont {Abiuso}, \citenamefont {Calsamiglia},\ and\ \citenamefont {Perarnau-Llobet}}]{Puig2025DynamicalSteadyStateManyBodyMetrology}%
  \BibitemOpen
  \bibfield  {author} {\bibinfo {author} {\bibfnamefont {R.}~\bibnamefont {Puig}}, \bibinfo {author} {\bibfnamefont {P.}~\bibnamefont {Sekatski}}, \bibinfo {author} {\bibfnamefont {P.~A.}\ \bibnamefont {Erdman}}, \bibinfo {author} {\bibfnamefont {P.}~\bibnamefont {Abiuso}}, \bibinfo {author} {\bibfnamefont {J.}~\bibnamefont {Calsamiglia}},\ and\ \bibinfo {author} {\bibfnamefont {M.}~\bibnamefont {Perarnau-Llobet}},\ }\bibfield  {title} {\bibinfo {title} {From dynamical to steady-state many-body metrology: Precision limits and their attainability with two-body interactions},\ }\href {https://doi.org/10.1103/PRXQuantum.6.030309} {\bibfield  {journal} {\bibinfo  {journal} {PRX Quantum}\ }\textbf {\bibinfo {volume} {6}},\ \bibinfo {pages} {030309} (\bibinfo {year} {2025})}\BibitemShut {NoStop}%
\bibitem [{\citenamefont {Lenar{\v{c}}i{\v{c}}}\ \emph {et~al.}(2020)\citenamefont {Lenar{\v{c}}i{\v{c}}}, \citenamefont {Alberton}, \citenamefont {Rosch},\ and\ \citenamefont {Altman}}]{lenarvcivc2020critical}%
  \BibitemOpen
  \bibfield  {author} {\bibinfo {author} {\bibfnamefont {Z.}~\bibnamefont {Lenar{\v{c}}i{\v{c}}}}, \bibinfo {author} {\bibfnamefont {O.}~\bibnamefont {Alberton}}, \bibinfo {author} {\bibfnamefont {A.}~\bibnamefont {Rosch}},\ and\ \bibinfo {author} {\bibfnamefont {E.}~\bibnamefont {Altman}},\ }\bibfield  {title} {\bibinfo {title} {Critical behavior near the many-body localization transition in driven open systems},\ }\href {https://doi.org/10.1103/PhysRevLett.125.116601} {\bibfield  {journal} {\bibinfo  {journal} {Phys. Rev. Lett.}\ }\textbf {\bibinfo {volume} {125}},\ \bibinfo {pages} {116601} (\bibinfo {year} {2020})}\BibitemShut {NoStop}%
\bibitem [{\citenamefont {D’Alessio}\ and\ \citenamefont {Rigol}(2014)}]{d2014long}%
  \BibitemOpen
  \bibfield  {author} {\bibinfo {author} {\bibfnamefont {L.}~\bibnamefont {D’Alessio}}\ and\ \bibinfo {author} {\bibfnamefont {M.}~\bibnamefont {Rigol}},\ }\bibfield  {title} {\bibinfo {title} {Long-time behavior of isolated periodically driven interacting lattice systems},\ }\href {https://doi.org/10.1103/PhysRevX.4.041048} {\bibfield  {journal} {\bibinfo  {journal} {Phys. Rev. X}\ }\textbf {\bibinfo {volume} {4}},\ \bibinfo {pages} {041048} (\bibinfo {year} {2014})}\BibitemShut {NoStop}%
\bibitem [{\citenamefont {Lazarides}\ \emph {et~al.}(2014)\citenamefont {Lazarides}, \citenamefont {Das},\ and\ \citenamefont {Moessner}}]{lazarides2014equilibrium}%
  \BibitemOpen
  \bibfield  {author} {\bibinfo {author} {\bibfnamefont {A.}~\bibnamefont {Lazarides}}, \bibinfo {author} {\bibfnamefont {A.}~\bibnamefont {Das}},\ and\ \bibinfo {author} {\bibfnamefont {R.}~\bibnamefont {Moessner}},\ }\bibfield  {title} {\bibinfo {title} {Equilibrium states of generic quantum systems subject to periodic driving},\ }\href {https://doi.org/10.1103/PhysRevE.90.012110} {\bibfield  {journal} {\bibinfo  {journal} {Phys. Rev. E}\ }\textbf {\bibinfo {volume} {90}},\ \bibinfo {pages} {012110} (\bibinfo {year} {2014})}\BibitemShut {NoStop}%
\bibitem [{\citenamefont {Hauschild}\ and\ \citenamefont {Pollmann}(2018)}]{tenpy}%
  \BibitemOpen
  \bibfield  {author} {\bibinfo {author} {\bibfnamefont {J.}~\bibnamefont {Hauschild}}\ and\ \bibinfo {author} {\bibfnamefont {F.}~\bibnamefont {Pollmann}},\ }\bibfield  {title} {\bibinfo {title} {Efficient numerical simulations with tensor networks: Tensor network python ({TeNPy})},\ }\href {https://doi.org/10.21468/SciPostPhysLectNotes.5} {\bibfield  {journal} {\bibinfo  {journal} {SciPost Phys. Lect. Notes}\ ,\ \bibinfo {pages} {5}} (\bibinfo {year} {2018})}\BibitemShut {NoStop}%
\bibitem [{\citenamefont {Mi}\ \emph {et~al.}(2022)\citenamefont {Mi}, \citenamefont {Ippoliti},\ and\ \citenamefont {Quintana}}]{Mi2022TimeCrystallineEigenstateOrder}%
  \BibitemOpen
  \bibfield  {author} {\bibinfo {author} {\bibfnamefont {X.}~\bibnamefont {Mi}}, \bibinfo {author} {\bibfnamefont {M.}~\bibnamefont {Ippoliti}},\ and\ \bibinfo {author} {\bibfnamefont {e.~a.}\ \bibnamefont {Quintana}},\ }\bibfield  {title} {\bibinfo {title} {Time-crystalline eigenstate order on a quantum processor},\ }\href {https://doi.org/10.1038/s41586-021-04257-w} {\bibfield  {journal} {\bibinfo  {journal} {Nature}\ }\textbf {\bibinfo {volume} {601}},\ \bibinfo {pages} {531} (\bibinfo {year} {2022})}\BibitemShut {NoStop}%
\end{thebibliography}
%

\end{document}